\documentclass[a4paper,11pt]{article}
\pdfoutput=1

\usepackage{jcappub}
\usepackage{bm}
\usepackage{placeins}
\usepackage{xcolor}
\usepackage{float}
\usepackage{booktabs}
\usepackage{makecell}
\usepackage{eso-pic}

\title{Semi-Analytic Predictions for Strong-Lensing Perturbers in Cold and Self-Interacting Dark Matter}

\author[a,c]{Jonathan Tekverk,}
\author[b,c]{Duncan K. Adams,}
\author[a]{Simon Birrer,}
\author[c]{Rouven Essig}
\author[d]{and Daniel Gilman}

\affiliation[a]{Department of Physics \& Astronomy, Stony Brook University,
Stony Brook, NY 11794, USA}
\affiliation[b]{Department of Physics and Astronomy, University of California,
Irvine, CA 92697, USA}
\affiliation[c]{C.N.~Yang Institute for Theoretical Physics, Stony Brook University, Stony Brook, NY 11794, USA}
\affiliation[d]{Department of Astronomy \& Astrophysics, University of Chicago, Chicago, IL 60637, USA}

\emailAdd{jonathan.tekverk@stonybrook.edu}
\emailAdd{duncan.adams@stonybrook.edu}
\emailAdd{simon.birrer@stonybrook.edu}
\emailAdd{rouven.essig@stonybrook.edu}
\emailAdd{gilmanda@uchicago.edu}

\abstract{Strong gravitational lensing is sensitive to both the abundance and internal structure of dark matter halos, providing a probe of gravothermal evolution in self-interacting dark matter (SIDM). We combine semi-analytic cold dark matter (CDM) subhalo histories with a model for gravothermal evolution in SIDM to efficiently generate CDM and SIDM subhalo populations across SIDM parameter space. We predict their projected enclosed masses and deflection-based projected density slopes and compare these properties with those of three perturbers inferred through gravitational imaging in PJ011646, SDSS J0946+1006, and JVAS B1938+666.
Under the fiducial SIDM model parameters $\sigma_0/m_{\chi}=147.1\rm~cm^2/g$ and $w=120\rm~km/s$, SIDM subhalos reproduce all three perturbers more frequently than CDM subhalos. For PJ011646, we find on average $\sim0.6$ cored SIDM subhalos per host halo that reproduce the inferred structure, compared with $\sim0.002$ in CDM, while core-collapsed subhalos are too steep. For SDSS J0946+1006, we find $\sim3$ core-collapsed and $\sim0.004$ cored SIDM subhalos per host, compared with none in 900 CDM realizations, although baryons, which we neglect, may be important for such massive subhalos. For ``detection $\mathcal{V}$'' in JVAS B1938+666, we find $\sim220$ core-collapsed and $\sim20$ cored SIDM subhalos per host within the $2\sigma$ posterior region, compared with $\sim0.2$ in CDM; the CDM candidates within the $5\sigma$ region require exceptionally early accretion, with a median infall redshift of $z_{\rm infall}\sim13$. In our framework, simultaneously reproducing all three perturbers requires a turnover velocity of approximately $w\gtrsim10^2~\rm km/s$, so that the massive $\sim5.6\times10^{10}\rm~M_\odot$ progenitors of SDSS J0946+1006 candidates can undergo gravothermal collapse, and is achieved for $\sigma_0/m_\chi\gtrsim10^2~\rm cm^2/g$ among the parameter sets considered. Our analysis demonstrates the utility of semi-analytic models in quantifying how frequently subhalos reproduce individual perturbers in different dark matter models. Because our framework is modular, its predictions can be readily updated as models of tidal stripping and late-stage gravothermal evolution in SIDM improve. 
}

\begin{document}

\AddToShipoutPictureBG*{%
  \AtPageUpperLeft{%
    \hspace{0.875\paperwidth}%
    \raisebox{-7\baselineskip}{%
      \makebox[0pt][r]{YITP-SB-2026-17}
}}}%
\maketitle
\raggedbottom

\section{Introduction}

The particle nature of dark matter remains an open question, despite the success of the standard cold dark matter (CDM) paradigm in describing large-scale structure, the cosmic microwave background, and galaxy clustering~\citep{planck_2018_results, DES_survey_2022, SDSS_BAO_2021}. In this paradigm, dark matter particles are cold and collisionless, and structure forms hierarchically: small dark matter halos collapse earlier and form larger halos through sequential mergers~\citep{lacey_cole_1993, white_1978, press_schechter_1974}. This leads to characteristic predictions for both the abundance and internal structure of dark matter halos, including a rapidly increasing abundance toward lower halo masses, a relationship between halo mass and concentration, and universal density profiles~\citep{NFW_1997, dutton_2014_mass_conc_einsasto, tinker_2008_shmf}. This process leaves behind a large population of bound subhalos inside galaxy- and group-scale host halos, with typical host masses of
$\sim10^{12}~\mathrm{M}_\odot$ and $\sim10^{13}~\mathrm{M}_\odot$, respectively~\citep{gao_2004_subhalos,han_2016_subhalo_mass_func}. For both field halos and subhalos with masses below $\sim10^{8}~\mathrm{M}_\odot$, inefficient gas cooling, reionization, and feedback strongly suppress gas retention and star formation, rendering these halos invisible to traditional luminous tracers~\citep{zavala_2019_cdm_review, gao_2004_subhalos, sawala_2016_low_subhalos, okamoto_2008_low_subhalos}. As a consequence, CDM has not been robustly tested on these mass scales, and alternative hypotheses for the particle nature of dark matter can modify both the abundance and internal structure of these invisible halos. For example, warm dark matter (WDM) can decouple from the primordial plasma while still relativistic, producing a free-streaming scale that suppresses small-scale density fluctuations, thereby reducing the abundance and concentrations of low-mass halos~\citep{bode_2001_wdm, schneider_2012_wdm, ludlow_2016_wdm}. In fuzzy dark matter (FDM), the wave-like behavior of ultralight particles produces central solitonic cores in dark matter halos while suppressing their abundance at low masses~\citep{Fuzzy_DM_sims}. In self-interacting dark matter (SIDM), the focus of this work, scattering between dark matter particles transports heat within halos. This initially produces lower-density cores, which can subsequently collapse and develop high central densities~\citep{Nadler_2025_concerto_sims}. Measuring the abundance and internal structure of dark matter halos below $\sim10^{8}~\mathrm{M}_\odot$ would therefore provide a direct test of structure formation across different dark matter models.

Strong gravitational lensing can probe the abundance and density profiles of both field halos and subhalos through their gravitational effects on the light of background sources, without requiring these halos themselves to host stars or gas. Halos projected near lensed images can perturb the lensing potential and produce localized distortions in observations~\citep{Vegetti_2023_strong_lens_dm_review}. Flux-ratio anomalies of multiply imaged quasars can be used to statistically constrain the abundance and internal structure of low-mass halos~\citep{Gilman_2019_wdm_chills,gilman_2021_sidm_lensing,gilman_2026_jwst5}. In addition, gravitational imaging of extended lensed arcs and Einstein rings can reveal individual halos and constrain their masses and density slopes~\cite{vegetti_2009_inference,Tajalli_2025_J0946_sidm}. With sufficiently high-resolution observations, these approaches can probe dark matter substructure down to mass scales of $\sim10^6~\mathrm{M_\odot}$~\citep{gilman_2022_subhalos,gilman_2020_mass_conc_rel,Despali_2025_cdm_unexpect,Powell_2025_dir_det_original}. Using gravitational imaging, several studies have reported candidate detections of individual dark matter perturbers in the strong-lens systems SDSS J0946+1006~\cite{Vegetti_2010_dir_det_original}, JVAS B1938+666~\cite{Vegetti_2012_dir_det_original,Powell_2025_dir_det_original}, SDP.81~\cite{Hezaveh_2016_dir_det_original}, SPT2147-50~\cite{Lange_2025_SPT_direct_detec}, and PJ011646~\cite{PJ011646_2026_dir_det_original}, although the robustness and interpretation of these detections vary across analyses and depend on the adopted lens and source models. Follow-up studies have argued that the inferred compactness of some of these perturbers may be difficult to reconcile with CDM~\citep{Vegetti_2026_cdm_unexpect, Despali_2025_cdm_unexpect, minor_2025_direct_detect, Tajalli_2025_J0946_sidm}, a question we investigate in this work. With sufficiently robust theoretical predictions for substructure, strong lensing measurements can test CDM structure formation and distinguish between dark matter models that predict different internal density structures for low-mass halos~\citep{Diaz_Rivero_2018,sengul_2022_inference,gilman_2021_sidm_lensing,kollman_2026_SL_steep_profiles,kong_2025_concerto}.

The self-interacting dark matter paradigm introduces non-gravitational interactions between dark matter particles and was proposed to address small-scale tensions between CDM predictions and observations of dwarf and low-surface-brightness galaxies~\citep{Spergel_2000_sidm_idea, Tulin_2018_sidm_review}. Unlike collisionless dark matter, SIDM allows for heat transport within halos through particle self-interactions, which can significantly alter their inner density structure---a process known as gravothermal evolution~\citep{lynden_1968_gravothermal,lynden_1980_gravothermal,Yang_2022_viscosity_cs_sims,outmezguine_2023_sidm_evol}. This tends to produce cored halos in its early stages, but can eventually lead to core collapse, characterized by a rapidly increasing central density, a shrinking core radius, and a steepening outer density profile~\citep{balberg_2002_core_collapse,koda_sidm_core_collapse_sims,Essig_2019,Nishikawa_2020_core_collapse,outmezguine_2023_sidm_evol,Adhikari_2025_sidm_astro_tests}. At late times, core collapse may even lead to black hole formation~\citep{balberg_2002_sidm_BHs,Pollack_2015_sidm_collapse_bhs,Jiang_2025_sidm_black_holes}. SIDM is especially relevant for strong-lensing analyses because gravothermal evolution can produce a greater diversity of subhalo density profiles than CDM, with core-collapsed halos generally producing stronger lensing perturbations than their CDM counterparts~\citep{Nadler_2023_sidm_sims,Nadler_2025_concerto_sims,mace_2026_cored_collap_lensing_power,yang_haibo_2021_sidm_lensing_effect,gilman_2021_sidm_lensing}. At the population level, a recent analysis of flux ratios from 29 quadruply imaged quasars constrained the abundance of core-collapsed halos, finding a preference for SIDM models with core collapse over CDM~\citep{gilman_2026_jwst5}. Due to the compactness of individual core-collapsed halos, SIDM has also been proposed as a possible explanation for inferred strong-lens perturbers whose projected masses and density slopes may be difficult to realize in CDM, including those in SDSS J0946+1006 and JVAS B1938+666~\citep{Enzi_2024_cdm_unexpect,kong_2025_concerto,Vegetti_2026_cdm_unexpect}.

High-resolution $N$-body simulations that include dark matter self-interactions and self-consistently follow halo merger histories, tidal evolution, and gravothermal evolution can make robust predictions for the (sub)halo populations inferred from strong-lensing data~\citep{Nadler_2023_sidm_sims, Nadler_2025_concerto_sims}. For example, \cite{kong_2025_concerto} identified strong-lens perturber candidates in the SIDM \textsc{Concerto}~\cite{Nadler_2025_concerto_sims} simulations based on their projected masses and density slopes. However, resolving gravothermal collapse is itself numerically challenging because the central region of the halo becomes progressively denser and more compact, requiring increasingly stringent mass, spatial, and temporal resolution. Studies have shown that core-collapse rates can be sensitive to particle number, gravitational softening, and timestep choices, with the required particle resolution generally exceeding that achieved for low-mass subhalos in simulations~\citep{sidm_sims_limits1_conv_tests,sidm_sims_limits2}. Moreover, cosmological SIDM $N$-body simulations are computationally expensive, so they typically follow only a few host halos for a limited set of SIDM parameter choices. However, subhalos with projected masses and density slopes matching those of an observed perturber can be rare among simulated subhalo populations, and quantifying how often they arise, and how this frequency depends on the SIDM model, requires large ensembles of subhalo populations across many host realizations and SIDM parameter choices.

Semi-analytic models provide a much faster complementary approach because they sample halo merger histories from models calibrated against cosmological simulations and use analytic prescriptions for halo evolution calibrated against idealized $N$-body simulations, thereby avoiding the computational cost of simulating individual dark matter particle trajectories~\citep{Benson_2012_galacticus,satgen_1,satgen_2}. This approach can efficiently generate large ensembles of CDM subhalo populations in group-scale hosts and track the accretion and tidal evolution histories of individual subhalos~\citep{gannon_2025_dm_lensing,du_2025_faster_sam}. When combined with models of gravothermal evolution in SIDM, these CDM subhalo histories can be mapped into SIDM counterparts using calibrated prescriptions for their structural evolution. For example, the model of~\cite{Yang_2024_parametric}, which is calibrated to SIDM $N$-body simulations, describes both the core-expansion phase and the onset of core collapse during gravothermal evolution~\citep{Yang_nadler_2023_sidm_param_mw, Yang_2022_viscosity_cs_sims}. In~\cite{sashimi_ando_2025}, this mapping was applied to semi-analytic CDM subhalo populations for Rutherford-like velocity-dependent cross sections, demonstrating that the fraction of core-collapsed subhalos peaks at a characteristic mass scale set by the SIDM parameters. This suggests that strong lensing may be sensitive to specific regions of SIDM parameter space, for which semi-analytic models provide a computationally efficient means of exploration.

Although previous studies have identified subhalos in CDM and SIDM $N$-body simulations with projected structures consistent with inferred perturbers~\cite{Minor_2021_J0946_concentration,Despali_2025_cdm_unexpect,kong_2025_concerto,zhang_haibo_2026_sidm_jvasb,haibo_2026_sidm_jvasb}, semi-analytic predictions for the abundance of such SIDM subhalos across many realizations of strong-lens hosts remain largely unexplored. In this work, we use the open-source semi-analytic model \textsc{SatGen}~\cite{satgen_1,satgen_2} to generate CDM subhalo realizations of host halos representative of observed strong-lens systems~\citep{Vegetti_2023_strong_lens_dm_review,SL_host_mass_measure_slacs}. We then apply the ``parametric'' model of~\cite{Yang_2024_parametric} to construct corresponding SIDM populations across SIDM parameter space. We evaluate the projected enclosed masses and deflection-based projected density slopes of CDM and SIDM subhalos and compare these properties with those of three perturbers inferred through gravitational imaging in the strong-lens systems PJ011646, SDSS J0946+1006, and JVAS B1938+666. This approach allows us to quantify how frequently CDM and SIDM perturber candidates arise, examine which evolutionary histories produce them, and test whether the inferred structure of multiple perturbers can be simultaneously reproduced with a single set of SIDM model parameters.

This paper is organized as follows. 
Section~\ref{sec:methodology} defines our methodology for generating CDM and SIDM subhalo populations and making lensing-relevant predictions. In Section~\ref{sec:results:theory}, we explore the dependence of the projected structure of CDM and SIDM subhalos across SIDM parameter space and various host halo redshifts. In Section~\ref{sec:results:perturbers}, we compare CDM and SIDM subhalos with three candidate strong-lens perturbers inferred through gravitational imaging in PJ011646, SDSS J0946+1006, and JVAS B1938+666, focusing on the projected structure, abundance, and evolutionary histories of perturber candidates across SIDM parameter space and under alternative tidal stripping prescriptions in SIDM. Section~\ref{sec:discussion} discusses the main modeling uncertainties in our analysis, and Section~\ref{sec:conclusion} summarizes our results and outlook. Appendix~\ref{app:ann_num_den_appendix} presents the annular number densities of CDM and SIDM subhalos for PJ011646 and JVAS B1938+666. Appendix~\ref{app:oren_gurian_compare} tests the robustness of our results to the choice of collapsed density profile. Appendix~\ref{app:satgen_pyhalo_comparison} compares the projected subhalo mass functions and distributions of averaged convergence for CDM subhalos generated using \textsc{SatGen}~\cite{satgen_1,satgen_2} and \textsc{pyHalo}~\citep{Gilman_2019_wdm_chills}. Appendix~\ref{app:observed_perturbers} summarizes the current observations and analyses of the three inferred perturbers considered in this work.

\section{Methodology}
\label{sec:methodology}

In this section, we describe how we generate CDM and SIDM subhalo populations, determine which SIDM subhalos undergo gravothermal collapse, and evaluate their lensing-relevant projected properties. We also describe how we convert the posteriors of three perturbers inferred through gravitational imaging in PJ011646, SDSS J0946+1006, and JVAS B1938+666 into the same projected properties used for comparison with CDM and SIDM subhalos in Section~\ref{sec:results:perturbers}. The detailed modeling of subhalo evolution in CDM and SIDM remains an active area of research. We make specific assumptions and choose specific prescriptions for subhalo evolution in both paradigms, and future work can modify our predictions. In Section~\ref{sec:satgen_real_in_cdm}, we describe the merger and tidal evolution histories of CDM subhalos generated using the \textsc{SatGen}~\cite{satgen_1,satgen_2} semi-analytic model. In Section~\ref{sec:satgen_real_in_sidm}, we define the framework for mapping these CDM subhalo histories into SIDM counterparts using a model for gravothermal evolution in SIDM, which is calibrated to $N$-body simulations~\citep{Yang_2024_parametric}. In Section~\ref{sec:subhalo_den_profiles}, we describe the density profiles assigned to CDM subhalos, which evolve only through tidal evolution, and to SIDM subhalos, which evolve through both tidal and gravothermal evolution. In Section~\ref{sec:pop_behavior_sidm_mapping}, we examine the dependence of gravothermal evolution and collapse on the SIDM model parameters and subhalo infall properties. In Section~\ref{sec:proj_lensing_quantities}, we describe the formalism for evaluating the projected enclosed masses and deflection-based projected density slopes of CDM and SIDM subhalos. Lastly, Section~\ref{sec:posteriors} defines the pivot radii at which these projected properties are evaluated for comparison between CDM and SIDM subhalos and the three strong-lens perturbers considered in this work. 

\subsection{CDM subhalo histories}
\label{sec:satgen_real_in_cdm}

We use the open-source code \textsc{SatGen}~\cite{satgen_1,satgen_2} to generate CDM subhalo realizations of $M_{\rm vir}(z_{\rm lens})=10^{13}\rm~M_\odot$ host halos representative of typical galaxy-galaxy strong lenses~\cite{SL_host_mass_measure_slacs,Vegetti_2023_strong_lens_dm_review}, including those of the lens systems PJ011646, SDSS J0946+1006, and JVAS B1938+666 for comparison between subhalos and inferred strong-lens perturbers in Section~\ref{sec:results:perturbers}. In this work, $M_{\rm vir}$ and $R_{\rm vir}$ denote the redshift-dependent virial mass and radius following the convention in~\cite{Bryan_Norman_1998}. \textsc{SatGen} generates halo merger trees using the extended Press--Schechter formalism~\cite{lacey_cole_1993,bond_eps_1991,Bower_eps_1991} and the Monte Carlo algorithm of~\cite{parkinson_2008}, with the calibration adopted by~\cite{benson_2017}. The host halo and pre-infall subhalos are modeled with Navarro--Frenk--White (NFW)~\cite{NFW_1997} density profiles, with concentrations determined from their individual merger histories~\citep{Zhao_2009_4_percent_infall}. At infall, subhalo orbital initial conditions are sampled from distributions based on cosmological simulations~\citep{Li_2020}. After infall, each subhalo is evolved by integrating its orbit in the host potential from infall to the lens redshift, while modeling the effects of dynamical friction and mass loss due to tidal stripping~\citep{green_2021}. Throughout this work, we use \emph{tidal stripping} to refer to the removal of bound mass, and \emph{tidal evolution} to refer to stripping together with the structural response of the bound remnant. If a subhalo's bound mass falls below $10^{-5}m_{\rm vir,infall}$ during its evolution, it is classified as tidally disrupted and its subsequent evolution is terminated. At the lens redshift, we only include subhalos that lie within the host virial radius and whose bound masses exceed the merger-tree resolution threshold. The dark-matter-only host and subhalos are modeled without baryonic physics such as a host disk potential or ram-pressure stripping of satellite gas. In Appendix~\ref{app:satgen_pyhalo_comparison}, we show that the projected subhalo mass functions and distributions of subhalo convergence are consistent between CDM subhalos generated using \textsc{SatGen} and \textsc{pyHalo}. We adopt the Planck 2018 cosmological parameters throughout our analysis~\citep{planck_2018_results}. In the following section, we use the resulting CDM mass-accretion, structural, and tidal-evolution histories as inputs to the SIDM mapping.

\subsection{SIDM gravothermal evolution}
\label{sec:satgen_real_in_sidm}

We map \textsc{SatGen} CDM subhalo populations into corresponding SIDM populations by applying the ``parametric'' model of gravothermal evolution in~\citep{Yang_2024_parametric}. This model assumes that the mass-accretion and tidal-stripping histories of SIDM halos approximately follow those of their CDM counterparts while self-interactions modify their internal structures. For certain velocity-dependent SIDM models, simulations show that the subhalo mass function is fairly consistent between CDM and SIDM~\citep{Yang_nadler_2023_sidm_param_mw,Nadler_2023_sidm_sims,turner_2021_sidm_sims}. However, this assumption can break down in regimes where scattering between host and subhalo particles induces additional mass loss~\citep{Zheng_2022_life_story,Nadler_2020_sidm_mass_loss,klemmer_2026_sidm_survival}, or where cored subhalos are more susceptible to tidal disruption, reducing their abundance~\cite{Errani_2022_survival,klemmer_2026_sidm_survival,Zheng_2022_life_story}. Nonetheless, this model allows computationally efficient exploration of the velocity-dependent SIDM parameter space without requiring a separate cosmological SIDM simulation for each parameter choice, subject to the limitations of the model. 

The model of~\cite{Yang_2024_parametric} approximates the gravothermal evolution of SIDM halos by applying calibrated corrections to the structural evolution histories of their CDM counterparts, $V_{\max,\mathrm{CDM}}(t)$ and $R_{\max,\mathrm{CDM}}(t)$, where $V_{\max}=\max_r\sqrt{GM(<r)/r}$ is the maximum circular velocity, which can be defined for any density profile, and $R_{\max}$ is the radius at which it occurs. For an NFW profile, these are related to the characteristic scale density $\rho_s$ and scale radius $r_s$ through $V_{\max}=1.648~r_s\sqrt{G\rho_s}$ and $R_{\max}=2.1626~r_s$. The model assumes that gravothermal evolution is approximately self-similar when time is normalized by the halo collapse timescale~\cite{Essig_2019, balberg_2002_core_collapse,Pollack_2015_sidm_collapse_bhs},
\begin{equation}
t_c(t)=\frac{150}{C}\frac{1}
{\left[\sigma_{\mathrm{eff}}(t)/m_\chi\right]\rho_{\mathrm{eff}}(t)r_{\mathrm{eff}}(t)}
\frac{1}{\sqrt{4\pi G\rho_{\mathrm{eff}}(t)}},
\label{eq:t_c_original}
\end{equation}
where $C=0.75$ is calibrated against $N$-body simulations~\cite{koda_sidm_core_collapse_sims,Nishikawa_2020_core_collapse}, and
\begin{align}
v_{\mathrm{eff}}(t) &= 0.64~V_{\max,\mathrm{CDM}}(t),
\label{eq:veff} \\
r_{\mathrm{eff}}(t) &= \frac{R_{\max,\mathrm{CDM}}(t)}{2.1626},
\label{eq:reff} \\
\rho_{\mathrm{eff}}(t) &= \frac{1}{G} \left[ \frac{V_{\max,\mathrm{CDM}}(t)}
{1.648~r_{\mathrm{eff}}(t)} \right]^2,
\label{eq:rhoeff}
\end{align}
where $v_{\mathrm{eff}}$ is the approximate one-dimensional halo velocity dispersion~\citep{Yang_2022_viscosity_cs_sims}, while $r_{\rm eff}$ and $\rho_{\rm eff}$ are the effective NFW structural parameters. The dependence of gravothermal evolution on the SIDM particle model enters through the halo collapse timescale and the effective self-interaction cross section, which we adopt from Eq.~(4.2) of~\cite{Yang_2022_viscosity_cs_sims},
\begin{equation}
\frac{\sigma_{\mathrm{eff}}(t)}{m_\chi}=\frac{1}{512~v_{\mathrm{eff}}(t)^8}\int d\ln v~v^8 \left[\frac{2}{3}\frac{\sigma_V(v)}{m_\chi}\right]
\exp\left[-\frac{v^2}{4v_{\mathrm{eff}}(t)^2}\right],
\label{eq:sigma_eff(v)}
\end{equation}
where the viscosity cross section is
\begin{equation}
\frac{\sigma_V(v)}{m_\chi}= 6~(\sigma_0/m_\chi)~ \frac{(2+x)\ln(1+x)-2x}{x^3},~~~
x\equiv \left(\frac{v}{w}\right)^2.
\label{eq:sigma(v)}
\end{equation}
Here, $v$ is the relative velocity between scattered dark matter particles, $\sigma_0/m_\chi$ is the cross-section normalization, and $w$ is the velocity scale controlling the turnover of the self-interaction cross section. In this work, we adopt the SIDM model parameters $\sigma_0/m_{\chi}=147.1~{\rm cm^2/g}$ and $w=120~{\rm km/s}$ from~\cite{Nadler_2023_sidm_sims} as our ``fiducial'' choice, but we will present several examples of different parameter choices. These parameters were chosen in~\citep{Nadler_2023_sidm_sims} to produce core-collapsed subhalos relevant for the perturber in SDSS J0946+1006 while simultaneously producing large cores in low-concentration isolated halos.

For each CDM subhalo predicted by \textsc{SatGen}, we reconstruct the evolution histories of its structural parameters, $V_{\max,\rm CDM}(t)$ and $R_{\max,\rm CDM}(t)$, from formation, through infall, and to the lens redshift of interest. After infall, these structural parameters evolve according to the tidal tracks described in~\cite{Green_2019_tidal_tracks}, which depend on the subhalo's bound mass fraction and infall concentration. Using Eqs.~\eqref{eq:t_c_original}--\eqref{eq:sigma(v)}, we compute the collapse-timescale history $t_c(t)$ of each subhalo and its corresponding gravothermal evolution history, characterized by the parameter
\begin{equation}
\tau(t)=\int_{t_f}^{t}\frac{dt'}{t_c(t')},
\label{eq:tau_integral}
\end{equation}
where we define the subhalo formation time $t_f$ as the snapshot at which its mass is closest to $4\%$ of its infall mass~\citep{Zhao_2009_4_percent_infall}. Since this integral increases monotonically, the model does not allow heating to reverse gravothermal evolution. \cite{raut_2026_extended_parametric} proposed a phenomenological extension of this model in which accretion-induced heating can delay core collapse, and showed that it better reproduces the evolution of some field halos in cosmological SIDM simulations, including a group-scale simulation with our fiducial SIDM parameters. Since subhalos stop accreting after infall, this effect mainly concerns their pre-infall evolution, and neglecting it may overestimate the abundance of collapsed subhalos, particularly among the massive progenitors of the SDSS J0946+1006 perturber candidates (Section~\ref{sec:J0946}).

Following~\cite{Yang_2024_parametric}, we map each subhalo's CDM structural parameter histories into corresponding SIDM histories using
\begin{equation}
V_{\rm max,SIDM}(t)=V_{\max,\mathrm{CDM}}(t_f)
+\int_{t_f}^{t} dt'~
\frac{dV_{\max,\mathrm{CDM}}(t')}{dt'}
+\int_{t_f}^{t} \frac{dt'}{t_c(t')}~
\frac{dV_{\max,\mathrm{Model}}(\tau')}{d\tau'},
\label{eq:yang_integral_vmax}
\end{equation}
\begin{equation}
R_{\rm max,SIDM}(t)=R_{\max,\mathrm{CDM}}(t_f)
+\int_{t_f}^{t} dt'~
\frac{dR_{\max,\mathrm{CDM}}(t')}{dt'}
+\int_{t_f}^{t} \frac{dt'}{t_c(t')}~
\frac{dR_{\max,\mathrm{Model}}(\tau')}{d\tau'}.
\label{eq:yang_integral_rmax}
\end{equation}
The functions $dV_{\max,\mathrm{Model}}/d\tau'$ and $dR_{\max,\mathrm{Model}}/d\tau'$ are the derivatives of the fitting functions in Eq.~(2.4) of the arXiv version of~\cite{Yang_2024_parametric}, where $\tau'\equiv \tau(t')$. These fitting functions are calibrated against idealized dark-matter-only $N$-body simulations with NFW initial conditions and dark matter self-interactions~\cite{Yang_2022_viscosity_cs_sims}, and the resulting model is subsequently validated against a cosmological zoom-in SIDM simulation containing both isolated halos and subhalos~\citep{Yang_nadler_2023_sidm_param_mw,Yang_2024_parametric}. The second terms in Eqs.~\eqref{eq:yang_integral_vmax} and~\eqref{eq:yang_integral_rmax} capture the CDM evolution of the subhalo, which includes mass accretion and tidal stripping, while the third terms capture the changes to $V_{\rm max,SIDM}(t)$ and $R_{\rm max,SIDM}(t)$ induced by gravothermal evolution. In~\cite{Yang_2024_parametric}, the model fitting functions were calibrated through $\tau(t)=1.0$, beyond which the SIDM evolution terms were dropped. However, \cite{Yang_nadler_2025_tau_1_1} validated the model through $\tau(t)=1.1$ using an extended cosmological SIDM simulation. We therefore drop the SIDM evolution terms for $\tau(t)>1.1$, while retaining the CDM evolution terms. We additionally adopt the minimum values $V_{\rm max,SIDM}=2.0~\rm km/s$ and $R_{\rm max,SIDM}=0.1~\rm kpc$ to avoid numerical artifacts, following~\cite{Yang_2024_parametric}. Using the predicted $V_{\rm max,SIDM}(t)$, $R_{\rm max,SIDM}(t)$, and $\tau(t)$ histories of each subhalo, we invert Eq.~(2.4) of~\cite{Yang_2024_parametric} to obtain its fictitious initial NFW structural parameters $V_{\rm max,0}(t)$ and $R_{\rm max,0}(t)$ (and hence $\rho_{s,0}(t)$ and $r_{s,0}(t)$). These are then used to determine the structural parameters $\rho_s(t)$, $r_s(t)$, and $r_c(t)$ of the cored SIDM density profile through Eq.~(2.3) of the arXiv version of~\cite{Yang_2024_parametric}, as discussed in the next section.

\subsection{Subhalo density profiles and tidal evolution}
\label{sec:subhalo_den_profiles}

At the lens redshift $z_{\rm lens}$, CDM subhalos are assigned density profiles based on their tidal evolution, while SIDM subhalos are assigned different density profiles based on their tidal and gravothermal evolution. After infall, the density profiles of \textsc{SatGen} CDM subhalos are modeled using the transfer function of~\cite{Green_2019_tidal_tracks}, which is calibrated against the \textsc{DASH} library of idealized $N$-body simulations~\citep{Ogiya_dash_2019}:
\begin{equation}
\label{eq:green_profile}
\rho_{\scriptscriptstyle\mathrm{CDM}}(r,t) = H\!\left(r,f_b(t),c_{\rm infall}\right)\rho_
{\scriptscriptstyle\mathrm{NFW}}(r,t_{\rm infall}),
\end{equation}
where the transfer function $H$ depends on each subhalo's instantaneous bound mass fraction $f_b(t)=m_{\rm bound}(t)/m_{\rm infall}$ and infall concentration $c_{\rm infall}$, capturing the response of the inner halo to tidal mass loss through revirialization as well as the truncation of the outer parts of the halo. SIDM subhalos with $\tau(z_{\rm lens})\leq1.1$ are assigned a cored profile:
\begin{equation}
\rho_{\scriptscriptstyle\mathrm{SIDM}}^
{\scriptscriptstyle\mathrm{Cored}}(r,t)=
\frac{\rho_s(t)}{\displaystyle
\frac{\left(r^4+r_c(t)^4\right)^{1/4}}{r_s(t)}\left(1+\frac{r}{r_s(t)}\right)^2}.
\label{eq:daneng_prof}
\end{equation}
The core radius $r_c(t)$, density normalization $\rho_s(t)$, and scale radius $r_s(t)$ are the $\tau(t)$-dependent structural parameters determined from Eq.~(2.3) in the arXiv version of~\cite{Yang_2024_parametric}, which capture both the core expansion and the onset of core collapse of the halo. Throughout this work, we use the term ``cored'' to denote all SIDM subhalos with $\tau(z_{\rm lens})\leq1.1$. The model of~\cite{Yang_2024_parametric} predicts that halos reach their ``maximal core'' at $\tau(t)\simeq0.18$, after which the core contracts while its density rises, albeit much more slowly than in the later core-collapse phase. It is worth mentioning that the cored density profile in Eq.~\eqref{eq:daneng_prof} slightly mismodels the density slope at $\tau(t)\sim1$ (see, e.g., Figure~1 of~\cite{Yang_2024_parametric}). Once core collapse begins, heat flows outward from the core, and the outer parts of the halo approach a self-similar density profile, i.e., $\rho(r)/\rho_s \approx (r/r_s)^{-2.2}$ for $r \gg r_{\rm c}$. Although the core continues to shrink and its density increases, with the exact profile in the innermost regions remaining uncertain, the self-similar outer region of the halo is relatively static~\cite{balberg_2002_core_collapse}. Consequently, the time elapsed since the onset of core collapse has little effect on lensing observables probing physical scales $r \gg r_c$. We model collapsed halos with $\tau(z_{\rm lens}) > 1.1$ using a separate collapsed profile calculated via kinetic theory~\citep{Slone_2026}, which is also validated with idealized SIDM $N$-body simulations. This profile corresponds to a halo at $\tau(t)\sim2$. It also has a constant-density core below a radius $0.01r_s$, and is approximately described outside this core by a power law, $\rho(r)\propto r^{-2.2}$. The enclosed mass of this collapsed profile agrees with that of the NFW profile to within approximately $10\%$ at the virial radius, with a weak dependence on concentration. Throughout this work, we refer to SIDM subhalos assigned the profile from~\citep{Slone_2026} as ``collapsed'' to distinguish them from the ``cored'' subhalos.

\begin{figure}[t]
    \centering
    \includegraphics{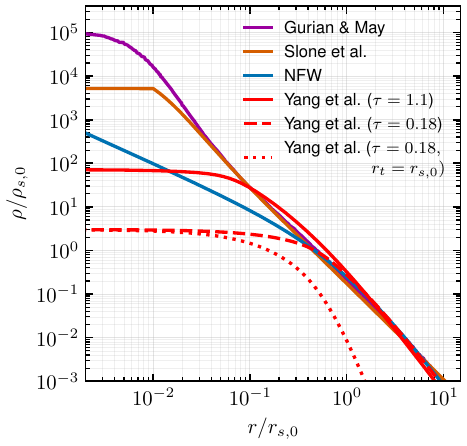}
    \caption{Density profiles assigned to SIDM subhalos, compared with the initial NFW profile (blue), with radius and density in units of the initial NFW scale radius $r_{s,0}$ and scale density $\rho_{s,0}$. Subhalos with $\tau(z_{\rm lens})\leq1.1$ are assigned the cored profile of Eq.~\eqref{eq:daneng_prof} (red), and those with $\tau(z_{\rm lens})>1.1$ the collapsed profile of~\cite{Slone_2026} (orange). The dashed red curve shows the cored profile at ``maximal core'', $\tau\simeq0.18$, and the solid red curve shows it at the collapse threshold, $\tau=1.1$, as predicted by the gravothermal evolution model of~\cite{Yang_2024_parametric}. The collapsed profile has a constant-density core inside $\approx0.01\,r_{s,0}$ and is approximately $\rho\propto r^{-2.2}$ outside it. The purple curve shows an alternative collapsed profile from Figure~1 of~\cite{gurian_may_deep_col} at $t/t_0=437$, where $t_0$ is the characteristic scattering timescale of the initial NFW halo; this corresponds to $\tau(t)\simeq12.2$ under the convention of Eq.~\eqref{eq:t_c_original}. Appendix~\ref{app:oren_gurian_compare} shows that our conclusions are robust to adopting this profile instead. All curves are shown without tidal truncation, except the dotted red curve, which shows the maximal-core profile truncated by the factor $(1+r/r_t)^{-5}$ with $r_t=r_{s,0}$.
    }
    \label{fig:density_profiles}
\end{figure}

Figure~\ref{fig:density_profiles} illustrates the density profiles assigned to SIDM subhalos under this binary prescription. For comparison, the purple curve shows a collapsed profile from~\cite{gurian_may_deep_col}, whose kinetic-theory simulations find departures from local thermodynamic equilibrium in the intermediate-mean-free-path regime that alter late-stage gravothermal evolution and the resulting density structure of collapsed halos relative to the conducting-fluid approximation. This profile reaches much higher central densities than our adopted collapsed profile, but the two are similar at radii beyond a few percent of $r_{s,0}$. In Appendix~\ref{app:oren_gurian_compare}, we show that our conclusions from Section~\ref{sec:results:perturbers} are robust to the choice of collapsed profile.

The cored and collapsed SIDM density profiles described above do not explicitly include tidal truncation or mass loss due to scattering between subhalo and host particles. The detailed modeling of subhalo tidal evolution, mass loss, and survival in SIDM remains an active area of research (see also the discussion in Section~\ref{sec:discussion}). We approximate tidal mass loss in SIDM by assuming that each SIDM subhalo retains the same fraction of its infall mass as its corresponding CDM counterpart. We implement this by applying a truncation factor of $(1+r/r_t(t))^{-5}$ to the density profiles of cored and collapsed subhalos, where a time-dependent truncation radius $r_t(t)$ is chosen to reproduce the bound mass fraction of the corresponding CDM subhalo at each time. The dotted red curve in Figure~\ref{fig:density_profiles} shows an example of a truncated cored SIDM profile. Figure~\ref{fig:pop_den} in Section~\ref{sec:results:perturbers} shows additional CDM and SIDM subhalo profiles that incorporate tidal evolution and, when relevant, gravothermal evolution.

\begin{figure}[t]
    \centering
    \includegraphics{
    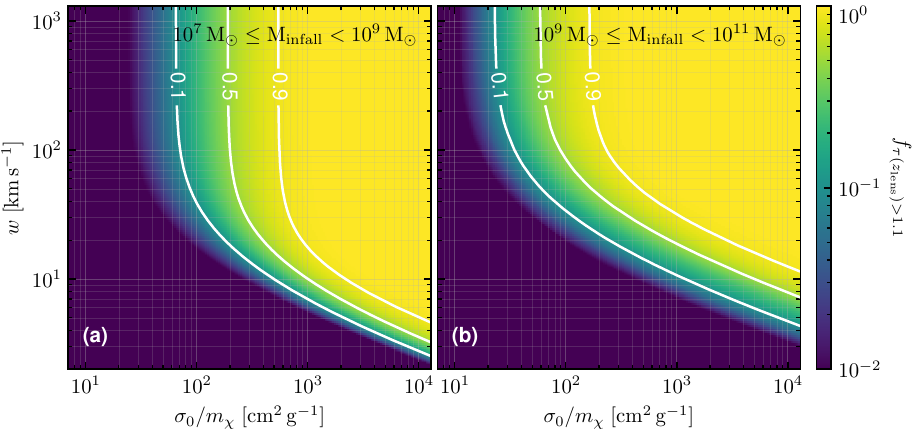}
    \caption{
    Fraction of gravothermally collapsed SIDM subhalos at the lens redshift $z_{\rm lens}=0.5$, i.e., those with $\tau(z_{\rm lens})>1.1$ as calculated using Eq.~\eqref{eq:tau_integral}, at each point in the $\left(\sigma_{0}/m_{\chi},w\right)$ parameter space (Eq.~\eqref{eq:sigma(v)}). \textbf{Panel (a):} collapse fractions $f_{\tau(z_{\rm lens})>1.1}$ of subhalos with infall virial masses $10^7~\mathrm{M}_\odot \le M_{\rm vir,infall}<10^9~\mathrm{M}_\odot$. \textbf{Panel (b):} corresponding collapse fractions for subhalos with $10^9~\mathrm{M}_\odot \le M_{\rm vir,infall}<10^{11}~\mathrm{M}_\odot$. We consider SIDM subhalos with bound masses $m_{\rm bound}>10^7\rm~M_\odot$ across 24 realizations of host halos with $M_{\rm vir}(z_{\rm lens})=10^{13}~\mathrm{M}_{\odot}$ at $z_{\rm lens}=0.5$. White contours mark constant collapse fractions of $0.1$, $0.5$, and $0.9$.
    }
    \label{fig:collapse-fraction-parameter-space}
\end{figure}

\subsection{Incidence of gravothermal collapse}
\label{sec:pop_behavior_sidm_mapping}

\begin{figure}[t]
    \centering
    \includegraphics{
    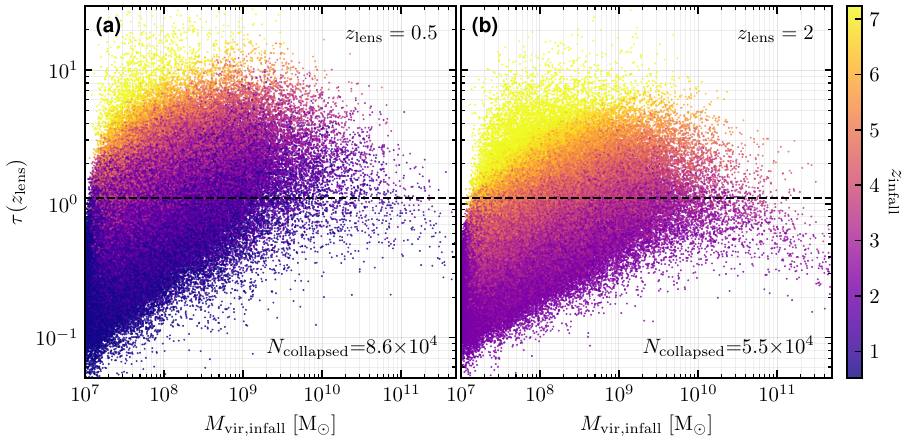}
    \caption{
    Gravothermal evolution parameters (Eq.~\eqref{eq:tau_integral}) of SIDM subhalos at the lens redshift $z_{\rm lens}$, under the fiducial SIDM model parameters $\sigma_0/m_\chi=147.1~{\rm cm^2/g}$ and $w=120~{\rm km/s}$. We consider SIDM subhalos with bound masses $m_{\rm bound}>10^7\rm~M_\odot$ across 24 realizations of host halos with $M_{\rm vir}(z_{\rm lens})=10^{13}~\mathrm{M}_{\odot}$ at each lens redshift. \textbf{Panel~(a):} $\tau(z_{\rm lens})$ versus subhalo infall virial mass $M_{\rm vir,infall}$ at $z_{\rm lens}=0.5$. Each point represents an individual subhalo, colored by its infall redshift $z_{\rm infall}$. The dashed horizontal line marks $\tau(z_{\rm lens})=1.1$, above which SIDM subhalos are classified as gravothermally collapsed. \textbf{Panel~(b):} $\tau(z_{\rm lens})$ versus $M_{\rm vir,infall}$ at $z_{\rm lens}=2$. There are fewer collapsed subhalos at $z_{\rm lens}=2$ than at $z_{\rm lens}=0.5$ because subhalos have comparatively less time for both tidal and gravothermal evolution.
    }
    \label{fig:tau_m_infall_z_infall}
\end{figure}

\begin{figure}[t]
    \centering
    \includegraphics{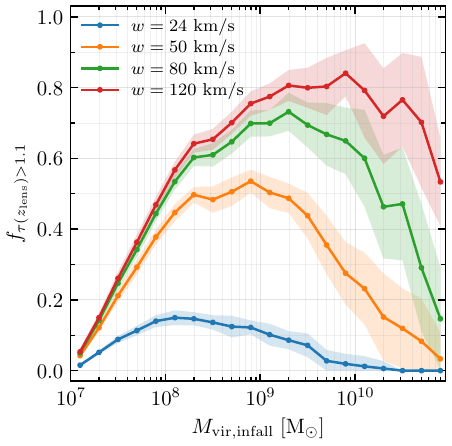}
    \caption{
    Fraction of gravothermally collapsed SIDM subhalos at the lens redshift $z_{\rm lens}=0.5$, i.e., those with $\tau(z_{\rm lens})>1.1$ as calculated using Eq.~\eqref{eq:tau_integral}, as a function of their infall virial masses. The cross-section normalization is fixed at $\sigma_{0}/m_{\chi}=147.1~{\rm cm^{2}~g^{-1}}$, while the turnover velocity is varied over $w=\{24,50, 80, 120\}$~km/s. Collapse fractions are evaluated in logarithmic infall mass bins of width 0.2~dex across 24 SIDM subhalo realizations of host halos with $M_{\rm vir}(z_{\rm lens})=10^{13}~\mathrm{M}_{\odot}$. The solid lines show the mean collapse fractions, while the shaded regions show the $1\sigma$ scatter across realizations. At fixed $\sigma_{0}/m_{\chi}$, the collapse fraction exhibits a characteristic peak as a function of infall mass that shifts toward larger masses as $w$ increases. Scatter in the CDM structural evolution histories broadens the infall mass range over which the collapse fraction peaks, because gravothermal evolution depends on the time-dependent effective cross section and halo structural parameters in Eq.~\eqref{eq:tau_int_history}.
    }
    \label{fig:col_frac_infall}
\end{figure}

In this subsection, we examine how subhalo infall virial mass and infall redshift correlate with gravothermal evolution and core collapse. We consider SIDM subhalos with bound masses $m_{\rm bound}>10^7\rm~M_\odot$ across 24 subhalo realizations of host halos with $M_{\rm vir}(z_{\rm lens})=10^{13}~\mathrm{M}_{\odot}$ at $z_{\rm lens}=0.5$. Figures~\ref{fig:collapse-fraction-parameter-space},~\ref{fig:tau_m_infall_z_infall}, and~\ref{fig:col_frac_infall} show how the incidence of gravothermal collapse depends on the SIDM model parameters and subhalo infall properties. Using Eqs.~\eqref{eq:t_c_original}--\eqref{eq:rhoeff}, Eq.~\eqref{eq:tau_integral} can be written as
\begin{equation}
\label{eq:tau_int_history}
\tau(t) \propto \int (\sigma_{\rm eff}(t')/m_\chi)~V_{\max,\rm CDM}^{3}(t')~R_{\max,\rm CDM}^{-2}(t')~dt',
\end{equation}
which shows how gravothermal evolution depends on the full CDM merger and tidal evolution histories of each subhalo through the time-dependent halo structural parameters $V_{\max,\rm CDM}(t)$ and $R_{\max,\rm CDM}(t)$.

Figure~\ref{fig:collapse-fraction-parameter-space} shows how the SIDM model parameters $\left(\sigma_{0}/m_{\chi},w\right)$ (Eq.~\eqref{eq:sigma(v)}) determine the fraction of gravothermally collapsed subhalos for two infall mass ranges,  with the underlying CDM subhalo histories held fixed. At fixed turnover velocity $w$, increasing the cross-section normalization $\sigma_{0}/m_{\chi}$ increases the effective cross section (Eq.~\eqref{eq:sigma_eff(v)}), shortens the halo collapse timescale (Eq.~\eqref{eq:t_c_original}), and drives a transition from negligible to nearly complete collapse in both panels. At smaller $w$, the self-interaction cross section is suppressed at higher velocities, which shifts contours of fixed collapse fraction toward larger $\sigma_0/m_\chi$, approximately following the high-velocity scaling $\sigma_{0}/m_{\chi}\propto w^{-4}$, whereas at sufficiently large $w$ the interaction approaches the velocity-independent limit and the contours become nearly vertical. The lower-mass subhalos in panel (a) typically have smaller internal velocities and can therefore undergo gravothermal collapse at comparatively smaller $w$ than the higher-mass subhalos in panel (b).

We now examine the gravothermal state of SIDM subhalos at the lens redshift under the fiducial SIDM model parameters $\sigma_0/m_\chi=147.1~{\rm cm^2/g}$ and $w=120~{\rm km/s}$. Figure~\ref{fig:tau_m_infall_z_infall} illustrates how subhalo infall mass and infall redshift correlate with $\tau(z_{\rm lens})$. We consider SIDM subhalos with bound masses $m_{\rm bound}>10^7\rm~M_\odot$ across 24 realizations of host halos with $M_{\rm vir}(z_{\rm lens})=10^{13}~\mathrm{M}_{\odot}$ at $z_{\rm lens}=0.5$ in panel~(a) and at $z_{\rm lens}=2$ in panel~(b). At fixed infall mass, subhalos that fall into the host earlier generally experience greater cumulative tidal stripping and reach larger values of $\tau(z_{\rm lens})$ due to accelerated gravothermal evolution. However, there are fewer collapsed subhalos at $z_{\rm lens}=2$ than at $z_{\rm lens}=0.5$ because subhalos have comparatively less time for both tidal and gravothermal evolution. The broad distributions of $\tau(z_{\rm lens})$ at fixed $M_{\rm vir,infall}$ or $z_{\rm infall}$ demonstrate that neither quantity uniquely determines the gravothermal state of SIDM subhalos; instead, gravothermal evolution depends on their full CDM structural evolution histories according to Eq.~\eqref{eq:tau_int_history}.

The relationship between the turnover velocity $w$ and the characteristic internal velocities of subhalos has important implications for gravothermal evolution. Figure~\ref{fig:col_frac_infall} shows how the fraction of SIDM subhalos that are gravothermally collapsed at $z_{\rm lens}=0.5$ varies with infall mass and turnover velocity $w$. We consider SIDM subhalos with bound masses $m_{\rm bound}>10^7\rm~M_\odot$ across 24 realizations of host halos with $M_{\rm vir}(z_{\rm lens})=10^{13}~\mathrm{M}_{\odot}$ at $z_{\rm lens}=0.5$. At fixed cross-section normalization $\sigma_0/m_\chi=147.1~{\rm cm^2/g}$, the collapse fraction exhibits a characteristic peak as a function of infall mass that shifts toward larger masses as the turnover velocity $w$ increases, consistent with the behavior found by~\cite{sashimi_ando_2025}. Increasing $w$ extends the range of internal halo velocities over which the effective cross section in Eq.~\eqref{eq:sigma_eff(v)} remains large, allowing progressively more massive subhalos to undergo gravothermal collapse. The location of this peak does not simply correspond to subhalos with $V_{\max}\simeq w$, for two reasons. First, the velocity-weighting in
Eq.~\eqref{eq:sigma_eff(v)} peaks at a relative velocity
$v=4v_{\rm eff}\simeq2.6\,V_{\max}$, so the effective
cross section becomes substantially suppressed when
$V_{\max}$ is of order $w/2.6$. Second, $\tau$ also depends on the structural factor $V_{\max}^3R_{\max}^{-2}$ in Eq.~\eqref{eq:tau_int_history}, which generally decreases toward lower masses, so low-mass subhalos evolve more slowly even where their cross section is not velocity-suppressed. The peak lies where these two effects balance, and because $V_{\max}$ and $R_{\max}$ evolve through accretion and tidal stripping, scatter in these histories broadens the mass range over which collapse is most frequent.

\subsection{Projected enclosed mass and deflection-based projected density slope}
\label{sec:proj_lensing_quantities}

Strong gravitational lensing is sensitive to the mass projected onto the plane of the sky rather than to the three-dimensional subhalo density profile. To connect with lensing observations, we therefore characterize subhalos by their projected enclosed masses and deflection-based projected density slopes, which correspond closely to quantities inferred through gravitational imaging~\citep{Despali_2025_cdm_unexpect}. Rather than projecting each three-dimensional density profile along the line of sight and subsequently integrating the surface mass density $\Sigma(R)=\int_{-\infty}^{\infty}\rho\big(\sqrt{R^2+z^2}\big)~dz$, where $z$ is the line-of-sight coordinate, we calculate the projected enclosed mass directly from the three-dimensional profile using a spherical-shell projection. A spherical shell at three-dimensional radius $r$ contributes fully to a cylinder of projected radius $R$ when $r\leq R$, and partially when $r>R$, giving
\begin{equation}
\label{eq:M2D_integral_weighted}
M_{\rm 2D}(<R)=4\pi \int_0^\infty \rho(r)~ W(R/r)~r^2dr ,
\end{equation}
where the projection weight is
\begin{equation}
\label{eq:kernel-cyl-sphere}
W(R/r)=
\begin{cases}
1, & r \le R, \\ 1 - \sqrt{1 - (R/r)^2},
& r > R .
\end{cases}
\end{equation}
For $r>R$, $W$ is the fraction of the shell's surface lying inside the cylinder, i.e.\ the two polar caps of half-angle $\theta_0=\arcsin(R/r)$, with combined area $4\pi r^2(1-\cos\theta_0)$. This reduces the calculation to a single one-dimensional integral.

We characterize the projected density profiles of circularly symmetric subhalos via the radial dependence of their deflection angle $\alpha(R)$, which, as for a point mass, depends only on the projected mass enclosed within $R$, $\alpha(R)\propto M_{\rm 2D}(<R)/R$~\citep{saha_2024_strong_lens_essentials}. For an exact power-law surface density, $\Sigma\propto R^{\Gamma_{\rm 2D}}$ (with $\Gamma_{\rm 2D}>-2$ so that the enclosed mass is finite), the enclosed mass and deflection angle scale as $M_{\rm 2D}(<R)\propto R^{\Gamma_{\rm 2D}+2}$ and $\alpha(R)\propto R^{\Gamma_{\rm 2D}+1}$, respectively. Therefore, we define the deflection-based projected density slope for an arbitrary projected subhalo profile as
\begin{equation}
\Gamma_{\rm 2D}(R)\equiv\frac{d\ln \alpha}{d\ln R}-1~.
\label{eq:gamma2d_deflection}
\end{equation}
More negative values of $\Gamma_{\rm 2D}$ correspond to steeper projected profiles. For reference, a three-dimensional power law $\rho\propto r^{-\gamma}$ with $1<\gamma<3$ projects to $\Sigma\propto R^{1-\gamma}$, so that $\Gamma_{\rm 2D}=1-\gamma$; an isothermal profile ($\gamma=2$) thus has $\Gamma_{\rm 2D}=-1$. For general profiles, this correspondence is only approximate, because $\Sigma(R)$ receives contributions from all $r\geq R$ along the line of sight. 

The deflection-based slope $\Gamma_{\rm 2D}$ also has simple physical bounds. Using $\alpha\propto M_{\rm 2D}/R$ and $dM_{\rm 2D}/dR=2\pi R~\Sigma(R)$, Eq.~\eqref{eq:gamma2d_deflection} can be written for any circularly symmetric profile as~\citep{saha_2024_strong_lens_essentials}
\begin{equation}
\Gamma_{\rm 2D}(R)=
\frac{d\ln M_{\rm 2D}(<R)}{d\ln R}-2\;=\;2\left[\frac{\Sigma(R)}{\bar\Sigma(<R)}-1\right],
\label{eq:gamma2d_mass}
\end{equation}
where $\bar\Sigma(<R)=M_{\rm 2D}(<R)/\pi R^2$. Thus, $\Gamma_{\rm 2D}$ compares the local surface density at $R$ with the mean surface density within $R$, rather than being the local logarithmic derivative of $\Sigma$. For the non-negative, monotonically decreasing density profiles considered in this work, the surface density $\Sigma(R)$ is also non-negative and decreasing, so $0\le\Sigma(R)\le\bar\Sigma(<R)$ and therefore 
\begin{equation}
-2\leq\Gamma_{\rm 2D}(R)\leq0~.
\label{eq:slope_bounds}
\end{equation}
The upper bound is approached where the surface density is nearly constant within $R$ (a core). The lower bound is the point-mass limit: when $R$ lies well outside most of the mass (e.g., beyond the truncation radius of a compact, tidally stripped subhalo), $\Sigma(R)\ll\bar\Sigma(<R)$, $M_{\rm 2D}$ approaches a constant, and $\alpha\propto R^{-1}$.

These limits provide a useful way to interpret the density profiles considered in this work. For all three subhalo families, tidal stripping removes mass along the line of sight at $r>R$, steepening the deflection-based slope toward the point-mass limit $\Gamma_{\rm 2D}\to-2$ as $R$ exceeds the truncation radius $r_t$. CDM subhalos (Eq.~\eqref{eq:green_profile}) retain an inner $\rho(r)\propto r^{-1}$ cusp, so $\Gamma_{\rm 2D}$ is relatively shallow when $R$ lies within the scale radius $r_s$. Before infall, an NFW profile has $\Gamma_{\rm 2D}\simeq-0.4$ at $R\simeq0.1r_s$ and $\simeq-0.9$ at $R\simeq r_s$. Cored SIDM subhalos (Eq.~\eqref{eq:daneng_prof}) approach the upper bound, $\Gamma_{\rm 2D}\to0$, for $R\ll r_c$. Collapsed subhalos mostly span $-2\lesssim\Gamma_{\rm 2D}\lesssim-1.25$: the shallow end is set by the untruncated profile at $r\gtrsim R$, corresponding to an effective three-dimensional slope $\gamma\simeq2.25$, with steeper slopes set by tidal truncation. In contrast to CDM and cored SIDM subhalos, this shallow end is nearly independent of subhalo mass, except for massive collapsed subhalos at small pivot radii, where $R$ approaches the $0.01r_s$ core radius of the collapsed profile; see Figure~\ref{fig:density_profiles} and Appendix~\ref{app:oren_gurian_compare}.

\subsection{Posteriors of inferred strong-lensing perturbers}
\label{sec:posteriors}

To compare strong-lens perturbers inferred in PJ011646, SDSS J0946+1006, and JVAS B1938+666 with simulated CDM and SIDM subhalo populations, we adopt the posterior samples from the corresponding gravitational-imaging analyses and convert them into posteriors of projected enclosed mass $M_{\rm 2D}(<R_{\rm pivot})$ and deflection-based projected density slope $\Gamma_{\rm 2D}(R_{\rm pivot})$ using Eqs.~\eqref{eq:M2D_integral_weighted} and~\eqref{eq:gamma2d_deflection}. $R_{\rm pivot}$ is the pivot radius at which the enclosed mass and density slope are most robustly constrained in each gravitational-imaging inference~\cite{Tajalli_2025_J0946_sidm,PJ011646_2026_dir_det_original}, and at which their covariance vanishes~\citep{tajalli_2026_pivot_radius}. It depends on the lensing geometry, perturber properties, signal-to-noise ratio, and angular resolution across each lens system. The pivot radius should not be confused with the perturbation radius of~\cite{minor_2025_direct_detect,Minor_2017_inference}, which is defined as the projected distance from the center of the perturber to the point on the critical curve that is most displaced by the perturber. The perturbation radius is set by the geometry of the perturbed critical curve, whereas the pivot radius is determined by where the inferred mass profile is best constrained, and the two need not coincide. For each CDM and SIDM density profile, we estimate $\Gamma_{\rm 2D}(R_{\rm pivot})$ by fitting the corresponding deflection profile over the radial window $0.75R_{\rm pivot}<R<1.25R_{\rm pivot}$, i.e.\ a half-width of $25\%$ of $R_{\rm pivot}$. This choice is motivated by previous lensing analyses of SDSS J0946+1006, which follow a similar convention~\cite{minor_2025_direct_detect,Enzi_2024_cdm_unexpect}, since the enclosed mass and density slope are constrained in a narrow radial range as opposed to a single radius. We use this window consistently when evaluating the deflection-based density slopes of all subhalos and perturbers considered in this work.

We summarize here the relevant details of each candidate perturber, leaving additional details to Appendix~\ref{app:observed_perturbers}. For the inferred perturber in PJ011646, we adopt the NFW posterior samples from~\cite{PJ011646_2026_dir_det_original}. Assuming that the perturber lies at the lens redshift $z_{\rm lens}=0.555$, the analysis reported $M_{200}=2.78^{+0.43}_{-0.66}\times10^{10}~\mathrm{M_\odot}$ and $c_{200}=30^{+5}_{-7}$, with a concentration that they argued is consistent with a typically tidally stripped CDM subhalo (see Section~\ref{sec:PJ011646} for our interpretation), and a projected enclosed mass $M_{\rm 2D}=3.57^{+0.16}_{-0.14}\times10^9~\mathrm{M_\odot}$ at an angular pivot radius of $0.29''$ (corresponding to a physical radius of $\simeq1.92~\mathrm{kpc}$ at $z_{\rm lens}=0.555$)~\cite{PJ011646_2026_dir_det_original}. When we convert the NFW posterior samples into a joint posterior of $M_{\rm 2D}(<1.92\rm~kpc)$ and $\Gamma_{\rm 2D}(1.92\rm~kpc)$, we obtain a median projected enclosed mass $M_{\rm 2D}(<1.92\rm~kpc)\simeq3.7\times 10^9~\rm M_{\odot}$ and a median deflection-based projected density slope $\Gamma_{\rm 2D}(1.92\rm~kpc)\simeq-0.9$. 

For the inferred perturber in SDSS J0946+1006 at lens redshift $z_{\rm lens}=0.222$, we adopt the truncated NFW posterior samples from~\cite{minor_2025_direct_detect}, whose lens model included both supersampling and multipoles; see their Table~4. At $95\%$ confidence, the analysis constrained the projected enclosed mass within $1~\mathrm{kpc}$ to $(2.2$--$3.4)\times10^9~\mathrm{M_\odot}$ and the average logarithmic slope of the local surface density, $d\mathrm{ln}\Sigma/d\mathrm{ln}R$, over $0.75$--$1.25~\mathrm{kpc}$ to be steeper than $-1.75$. This differs from the deflection-based slope and radial window adopted in this paper. From the truncated NFW posterior samples, we determine a pivot radius of $R_{\rm pivot}\simeq0.63~\mathrm{kpc}$, at which the fractional uncertainty in the projected enclosed mass is minimized and its covariance with the slope of the local surface density vanishes, following~\cite{tajalli_2026_pivot_radius}. We convert these posterior samples into a joint posterior of $M_{\rm 2D}(<0.63\rm~kpc)$ and $\Gamma_{\rm 2D}(0.63\rm~kpc)$, obtaining a median projected enclosed mass of $M_{\rm 2D}(<0.63\rm~kpc)\simeq2.2\times 10^9~\rm M_{\odot}$ and a median deflection-based projected density slope of $\Gamma_{\rm 2D}(0.63\rm~kpc)\simeq-1.5$. 

For ``detection $\mathcal{V}$'' in JVAS B1938+666 at $z_{\rm lens}=0.881$, we use the single-power-law (PL) inference reported in~\cite{Vegetti_2026_cdm_unexpect}, which gives $M_{\rm 2D}(<90~{\rm pc})=(1.182\pm0.033)\times10^6~\mathrm{M_\odot}$ and a three-dimensional logarithmic density slope $\gamma=2.330\pm0.032$. We adopt this posterior because its pure power-law form allows the inferred three-dimensional slope to be directly related to the deflection-based projected density slope, $\Gamma_{\rm 2D}(90\rm~pc)=1-\gamma\simeq-1.33$. As shown in Supplementary Figure~12 of~\cite{Vegetti_2026_cdm_unexpect}, the joint posterior of $M_{\rm 2D}(<90~\mathrm{pc})$ and $\Gamma_{\rm 2D}$ is approximately Gaussian, with weak correlation between the two quantities. However, the highest-evidence model consists of a central point mass with a disk of uniform surface mass density. The data robustly constrain $M_{\rm 2D}(<90~\mathrm{pc})\simeq1.2\times10^6~\mathrm{M_\odot}$ under both models and many alternative models at the lens redshift, despite their differing Bayesian evidences; see their Table~1~\citep{Vegetti_2026_cdm_unexpect}. Although~\cite{Powell_2025_dir_det_original} identified 80~pc as the approximate radius at which the uncertainty in the enclosed mass is minimized for their pseudo-Jaffe model, we adopt $R_{\rm pivot}=90\rm~pc$ following~\cite{Vegetti_2026_cdm_unexpect}, who found that 10 of the 12 models with $\Delta\ln E>-40$ predict $M_{\rm 2D}(<90\rm~pc)$ within 3\% of their highest-evidence model. In our work, all conversions between angular and physical scales are performed using the gravitational-lensing software \textsc{lenstronomy}~\citep{birrer_2018_Lens1,Birrer_2021_Lens2}, which internally uses \textsc{Astropy} for cosmological distance calculations~\citep{astropy3}.

\section{Projected structure of CDM and SIDM subhalos}
\label{sec:results:theory}

Before comparing with specific inferred perturbers in Section~\ref{sec:results:perturbers}, we first examine the general predictions of our framework for the projected structure of CDM and SIDM subhalos. In Section~\ref{sec:theory_varying_sidm_param}, we show how the distributions of projected enclosed masses $M_{\rm 2D}(<R_{\rm pivot})$ and deflection-based projected density slopes $\Gamma_{\rm 2D}(R_{\rm pivot})$ change across SIDM parameter space, and in Section~\ref{sec:theory_varying_z_pivot}, we examine their dependence on lens redshift and pivot radius.

\subsection{Dependence across SIDM parameter space}
\label{sec:theory_varying_sidm_param}

\begin{figure}[p]
    \centering
    \includegraphics{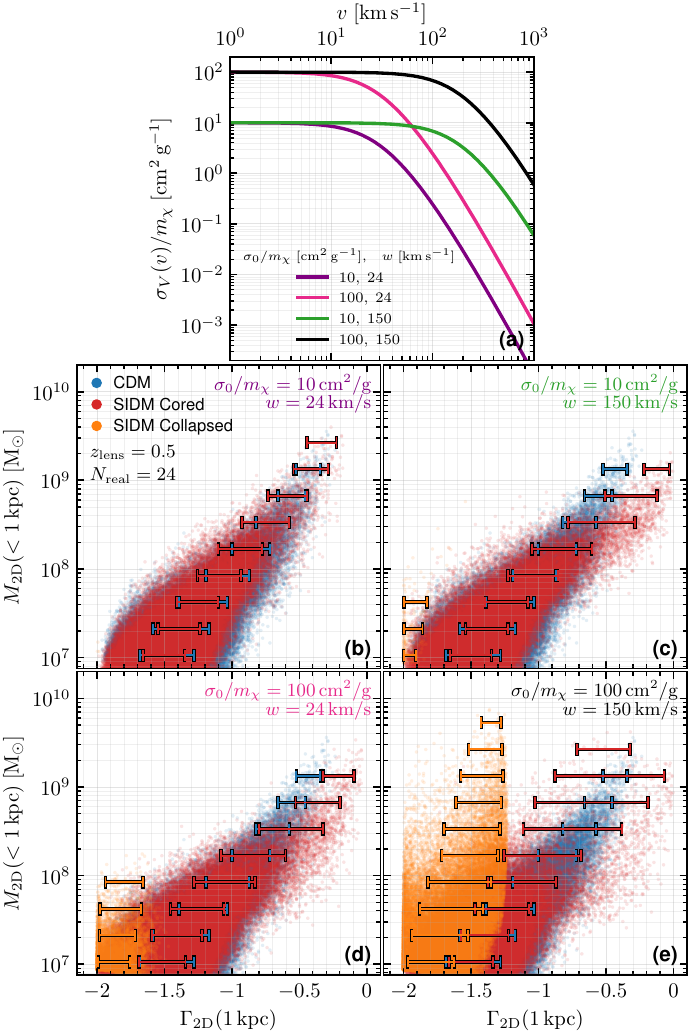}
    \caption{
    Dependence of the projected enclosed masses and deflection-based projected density slopes of CDM and SIDM subhalos on the SIDM model parameters $(\sigma_0/m_\chi,w)$ at fixed lens redshift $z_{\rm lens}=0.5$ and pivot radius $R_{\rm pivot}=1\rm~kpc$. \textbf{Panel~(a):} velocity-dependent viscosity cross section per unit mass $\sigma_V(v)/m_\chi$ (Eq.~\eqref{eq:sigma(v)}) for four sets of SIDM model parameters. \textbf{Panels~(b)--(e):} $M_{\rm 2D}(<R_{\rm pivot})$ (Eq.~\eqref{eq:M2D_integral_weighted}) versus $\Gamma_{\rm 2D}(R_{\rm pivot})$ (Eq.~\eqref{eq:gamma2d_deflection}), obtained by fitting a power law to the deflection profile over $0.75R_{\rm pivot}<R<1.25R_{\rm pivot}$. We consider CDM and SIDM subhalos with bound masses $m_{\rm bound}>10^7\rm~M_\odot$ across 24 realizations of host halos with $M_{\rm vir}(z_{\rm lens})=10^{13}~\mathrm{M}_{\odot}$. The SIDM model parameters are indicated in the upper-right corner of each panel, with colors matching the corresponding curves in panel~(a). Blue, red, and orange points denote CDM, cored SIDM, and collapsed SIDM subhalos, respectively. Horizontal bars denote the central $68\%$ interval of $\Gamma_{\rm 2D}$ in bins of $M_{\rm 2D}$ of width $0.3$~dex. CDM and SIDM subhalos can be distinguished in observations provided that the pivot radius probes a region where their density profiles differ, rather than lying outside their truncation radii.
    }
    \label{fig:proj_struc_fig}
\end{figure}

In Figure~\ref{fig:proj_struc_fig}, we show the distributions of projected enclosed mass and deflection-based projected density slope, at a fixed pivot radius $R_{\rm pivot}=1\rm~kpc$ using Eqs.~\eqref{eq:M2D_integral_weighted} and~\eqref{eq:gamma2d_deflection}, for 24 CDM and SIDM subhalo realizations at $z_{\rm lens}=0.5$ and four sets of SIDM model parameters. The CDM and SIDM populations exhibit a positive correlation between $M_{\rm 2D}$ and $\Gamma_{\rm 2D}$. At fixed pivot radius, more massive subhalos are generally probed farther within their characteristic structure and therefore have shallower projected slopes, whereas lower-mass or more strongly stripped subhalos are generally probed in their outer regions and tend to have steeper slopes. This trend is visible in the shift of the $68\%$ intervals toward shallower density slopes with increasing enclosed mass, with the CDM and SIDM populations overlapping in parts of the $M_{\rm 2D}$--$\Gamma_{\rm 2D}$ plane. Scatter in this correlation reflects differences in progenitor mass, concentration, and tidal evolution. 

CDM and SIDM subhalos with comparable $M_{\rm 2D}$ can be distinguished if their $\Gamma_{\rm 2D}$ are steeper or shallower. Panel (a) considers four velocity-dependent viscosity cross sections (Eq.~\eqref{eq:sigma(v)}), with colors corresponding to the parameter choices in the top right of panels (b)--(e). We consider subhalos with bound masses $m_{\rm bound}>10^7\rm~M_\odot$ across 24 realizations of host halos with $M_{\rm vir}(z_{\rm lens})=10^{13}\rm~M_{\odot}$ at $z_{\rm lens}=0.5$. For $\sigma_0/m_{\chi}=10~\rm cm^2/g$ and $w=24~\rm km/s$, panel (b) shows that the cored SIDM populations are largely degenerate with their CDM counterparts with comparable $M_{\rm 2D}$ because inefficient self-scattering produces core radii that are typically much smaller than the pivot radius, yielding comparable $\Gamma_{\rm 2D}$. When the turnover velocity is increased to $w=150~\rm km/s$ for fixed $\sigma_0/m_{\chi}=10~\rm cm^2/g$ in panel (c), $\sim0.1\%$ of SIDM subhalos are collapsed, predominantly at lower masses, while the most massive cored SIDM subhalos become shallower than their CDM counterparts because their core radii are closer to the pivot radius. Similar behavior is seen in panel (d), which uses a much larger cross-section normalization of $\sigma_0/m_{\chi}=100~\rm cm^2/g$ with $w=24~\rm km/s$. In this case, $\sim3\%$ of SIDM subhalos are collapsed while the most massive subhalos form cores comparable to the pivot radius because the cross section remains sufficiently large at higher internal halo velocities, as illustrated by the magenta curve in panel (a). When both the cross-section normalization and turnover velocity are increased to $\sigma_0/m_{\chi}=100~\rm cm^2/g$ and $w=150~\rm km/s$ in panel (e), the most massive cored subhalos exhibit both steeper and shallower slopes at the pivot radius compared to their CDM counterparts, while the lower-mass cored subhalos remain largely degenerate with their CDM counterparts because the pivot radius generally probes outside of their truncation radii. In addition, $\sim22\%$ of SIDM subhalos are collapsed, including higher-mass subhalos, since the self-scattering cross section remains high at larger internal halo velocities. For sufficiently large cross-section normalizations and turnover velocities, gravothermal evolution produces SIDM subhalos with $\Gamma_{\rm 2D}$ that can be substantially steeper or shallower than those of CDM subhalos with comparable $M_{\rm 2D}$. These subhalos can be distinguished in observations provided that the pivot radius probes a region where their density profiles differ, rather than lying outside their truncation radii.

\subsection{Dependence on pivot radius and redshift}
\label{sec:theory_varying_z_pivot}

\begin{figure}[p]
    \centering
    \includegraphics{
    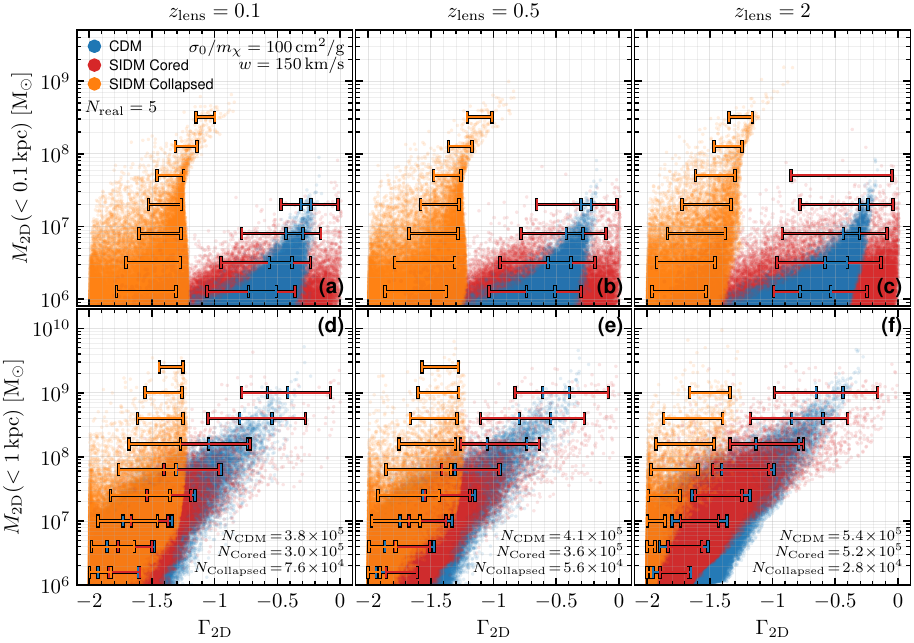}
    \caption{
    Projected enclosed masses $M_{\rm 2D}(<R_{\rm pivot})$ (Eq.~\eqref{eq:M2D_integral_weighted}) and deflection-based projected density slopes $\Gamma_{\rm 2D}(R_{\rm pivot})$ (Eq.~\eqref{eq:gamma2d_deflection}) of CDM and SIDM subhalos for different lens redshifts $z_{\rm lens}$ and pivot radii $R_{\rm pivot}$, obtained by fitting a power law to the deflection profile over $0.75R_{\rm pivot}<R<1.25R_{\rm pivot}$. Blue, red, and orange points denote CDM, cored SIDM, and collapsed SIDM subhalos, respectively, under the SIDM model parameters $\sigma_0/m_{\chi}=100\rm~cm^2/g$ and $w=150\rm~km/s$. In each panel, we consider CDM and SIDM subhalos with bound masses $m_{\rm bound}>10^6\rm~M_\odot$ across 5 realizations of host halos with $M_{\rm vir}(z_{\rm lens})=10^{13}~\mathrm{M}_{\odot}$ at each lens redshift. \textbf{Top panels:} $M_{\rm 2D}(<0.1\rm~kpc)$ versus $\Gamma_{\rm 2D}(0.1\rm~kpc)$ for lens redshifts $z_{\rm lens}=\{0.1,0.5,2\}$ from left to right. \textbf{Bottom panels:} $M_{\rm 2D}(<1\rm~kpc)$ versus $\Gamma_{\rm 2D}(1\rm~kpc)$ for the same lens redshifts. Horizontal bars denote the central $68\%$ interval of $\Gamma_{\rm 2D}$ in bins of $M_{\rm 2D}$ of width $0.4$~dex. For each lens redshift, the bottom panel gives the total numbers of CDM, cored SIDM, and collapsed SIDM subhalos with $m_{\rm bound}>10^6~\mathrm{M}_\odot$ across the 5 realizations. The number of CDM subhalos with bound masses $m_{\rm bound}>10^6\rm~M_\odot$ increases with lens redshift, since subhalos at higher redshifts have had less time to undergo tidal disruption after accretion into their hosts than those at lower redshift, consistent with theoretical expectations~\citep{jiang_2016_shmf_vs_z}. In contrast, there are fewer collapsed subhalos at higher redshifts because subhalos have had less time to undergo tidal and gravothermal evolution.
    }
    \label{fig:proj_struc_fig2}
\end{figure}

In Figure~\ref{fig:proj_struc_fig2}, we show the distributions of projected enclosed masses $M_{\rm 2D}(<R_{\rm pivot})$ and deflection-based projected density slopes $\Gamma_{\rm 2D}(R_{\rm pivot})$ of CDM and SIDM subhalos under the SIDM model parameters $\sigma_0/m_{\chi}=100~\rm cm^2/g$ and $w=150~\rm km/s$, evaluated at two pivot radii $R_{\rm pivot}=\{0.1\rm~kpc,1~kpc\}$ (rows) and three lens redshifts $z_{\rm lens}=\{0.1,0.5,2\}$ (columns). We choose these redshifts because galaxy-scale strong lenses are observed over a broad redshift range, from $z_{\rm lens}\sim0.03$ in nearby spectroscopic samples~\cite{smith_z0_lenses_2015} to $z_{\rm lens}\sim2$ in recent JWST discoveries~\citep{mercier_z2_lens_cosmos,nightingale_2025_z2_lens_discover}. At each lens redshift and pivot radius, we consider CDM and SIDM subhalos with bound masses $m_{\rm bound}>10^6\rm~M_\odot$ across 5 realizations of host halos with $M_{\rm vir}(z_{\rm lens})=10^{13}\rm~M_{\odot}$. At fixed pivot radius and SIDM model parameters, the CDM and SIDM subhalos span approximately the same regions in the $M_{\rm 2D}-\Gamma_{\rm 2D}$ plane over the range of lens redshifts considered. However, the distributions of $M_{\rm 2D}$ and $\Gamma_{\rm 2D}$ depend strongly on the pivot radius, which determines the physical scale over which the subhalo density profiles are probed. At $R_{\rm pivot}=0.1$~kpc, CDM and cored SIDM subhalos generally have shallower slopes than at $1$~kpc because the smaller pivot radius probes regions well inside their scale radii and, for cored subhalos, inside their cores. Collapsed subhalos instead remain steep at both pivot radii, except for the most massive ones at $R_{\rm pivot}=0.1$~kpc, where the pivot radius approaches the small core of the collapsed profile and their slopes become shallower (Section~\ref{sec:proj_lensing_quantities}). At each lens redshift, every CDM subhalo has either a cored or collapsed SIDM counterpart, and therefore the numbers of CDM and SIDM subhalos are equal. The number of CDM subhalos with bound masses $m_{\rm bound}>10^6\rm~M_\odot$ increases with lens redshift, since subhalos at higher redshifts have had less time to undergo tidal disruption after accretion into their hosts than those at lower redshift, consistent with theoretical expectations~\citep{jiang_2016_shmf_vs_z}. In contrast, there are fewer collapsed subhalos at higher redshifts because subhalos have had less time to undergo tidal and gravothermal evolution.

\section{Confronting strong-lensing perturbers with CDM and SIDM subhalos}
\label{sec:results:perturbers}

In this section, we compare the projected enclosed masses $M_{\rm 2D}(<R_{\rm pivot})$ (Eq.~\eqref{eq:M2D_integral_weighted}) and deflection-based projected density slopes $\Gamma_{\rm 2D}(R_{\rm pivot})$ (Eq.~\eqref{eq:gamma2d_deflection}) of CDM and SIDM subhalo populations with those of three strong-lens perturbers inferred in PJ011646~\cite{PJ011646_2026_dir_det_original}, SDSS J0946+1006~\cite{minor_2025_direct_detect}, and JVAS B1938+666~\cite{Vegetti_2026_cdm_unexpect}. For each system, we evaluate these properties at the pivot radius $R_{\rm pivot}$ where they are most robustly constrained in each gravitational-imaging inference. The perturbers in PJ011646, SDSS J0946+1006, and JVAS B1938+666 were inferred from submillimeter (ALMA), optical (HST), and radio (VLBI) observations, respectively. Differences in the lensing geometry, perturber properties, signal-to-noise ratio, and angular resolution across the three systems lead to pivot radii ranging from $1.92$~kpc to $90$~pc (Section~\ref{sec:posteriors}). In Sections~\ref{sec:PJ011646}--\ref{sec:JVASB_analogs}, we determine the numbers of CDM, cored SIDM, and collapsed SIDM subhalos across many realizations whose projected structures are approximately consistent with those of the inferred perturbers, i.e., those that approximately lie within the posterior regions in the $M_{\rm 2D}$--$\Gamma_{\rm 2D}$ plane, under the fiducial SIDM model parameters $\sigma_0/m_{\chi}=147.1\rm~cm^2/g$ and $w=120\rm~km/s$. We examine the evolutionary properties of these perturber candidates, focusing on their infall masses and bound mass fractions. In Section~\ref{sec:vary_sidm_param}, we test whether reproducing the inferred structures of all three perturbers requires specific choices of the SIDM model parameters, while in Section~\ref{sec:vary_sidm_strip}, we test the robustness of our results to alternative tidal stripping prescriptions for SIDM subhalos.

\begin{figure}[p]
    \centering
    \includegraphics{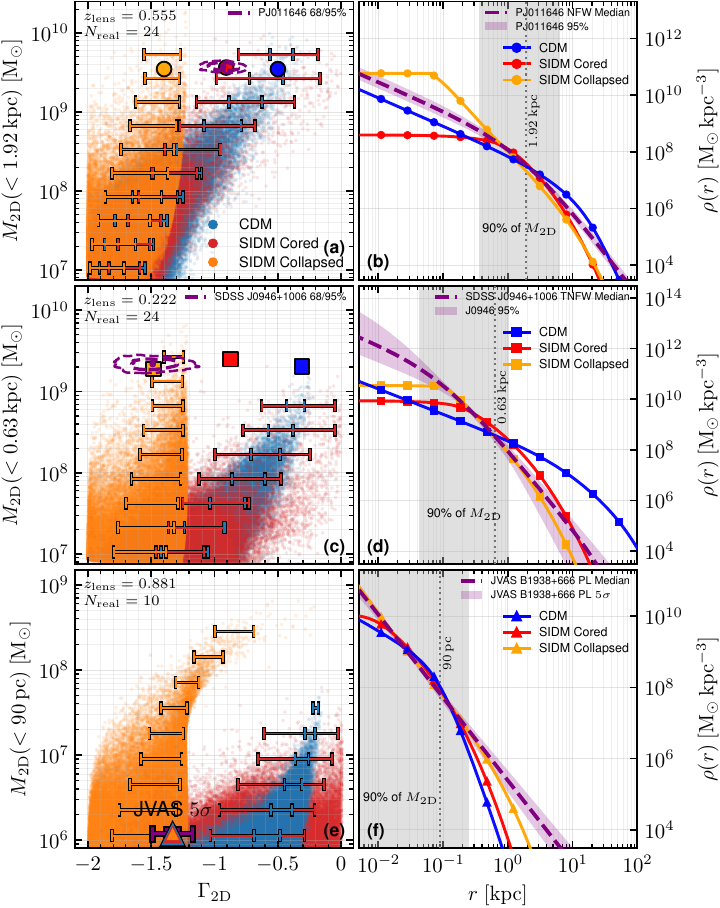}
    \caption{
    Comparison of the projected structures of three inferred strong-lens perturbers with those of CDM and SIDM subhalo populations under the fiducial SIDM model parameters $\sigma_0/m_{\chi}=147.1\rm~cm^2/g$ and $w=120\rm~km/s$. \textbf{Left panels:} projected enclosed mass $M_{\rm 2D}(<R_{\rm pivot})$ (Eq.~\eqref{eq:M2D_integral_weighted}) versus deflection-based projected density slope $\Gamma_{\rm 2D}(R_{\rm pivot})$ (Eq.~\eqref{eq:gamma2d_deflection}), obtained by fitting a power law to the deflection profile over $0.75R_{\rm pivot}<R<1.25R_{\rm pivot}$. Blue, red, and orange points denote CDM, cored SIDM, and collapsed SIDM subhalos, respectively, across $N_{\rm real}$ realizations. Horizontal bars show the central $68\%$ interval of $\Gamma_{\rm 2D}$ in bins of $M_{\rm 2D}$ of width $0.3$~dex. Purple contours show the $68\%$ and $95\%$ posterior regions of the perturbers in PJ011646 and SDSS J0946+1006; for ``detection $\mathcal{V}$'' in JVAS B1938+666, the purple marker shows the $5\sigma$ region. \textbf{Top row:} PJ011646, $z_{\rm lens}=0.555$, $R_{\rm pivot}=1.92$~kpc, $N_{\rm real}=24$. \textbf{Middle row:} SDSS J0946+1006, $z_{\rm lens}=0.222$, $R_{\rm pivot}=0.63$~kpc, $N_{\rm real}=24$. \textbf{Bottom row:} JVAS B1938+666, $z_{\rm lens}=0.881$, $R_{\rm pivot}=90$~pc, $N_{\rm real}=10$. \textbf{Right panels:} posterior-median density profiles (dashed purple) with their $95\%$ shaded regions ($5\sigma$ for JVAS B1938+666), compared with the profiles of three individual subhalos marked by circles, squares, or triangles in the corresponding left panels. The dotted vertical line marks $R_{\rm pivot}$. The gray band shows the range of three-dimensional radii over which spherical shells contribute to the central $90\%$ of $M_{\rm 2D}(<R_{\rm pivot})$ for the posterior-median profile, i.e.\ the radii to which the projected-mass constraint is most sensitive; it extends beyond $R_{\rm pivot}$ because shells at $r>R_{\rm pivot}$ also contribute along the line of sight.
    }
    \label{fig:pop_den}
\end{figure}

The left panels of Figure~\ref{fig:pop_den} show $M_{\rm 2D}(<R_{\rm pivot})$ versus $\Gamma_{\rm 2D}(R_{\rm pivot})$ for the three inferred strong-lens perturbers and the corresponding CDM and SIDM subhalo populations, under the fiducial SIDM model parameters $\sigma_0/m_{\chi}=147.1\rm~cm^2/g$ and $w=120\rm~km/s$. The deflection-based projected density slope $\Gamma_{\rm 2D}(R_{\rm pivot})$ is estimated by fitting a power law to the deflection profile, $\alpha\propto R^{\Gamma_{\rm 2D}+1}$, over $0.75R_{\rm pivot}<R<1.25R_{\rm pivot}$. We consider subhalos with bound masses $m_{\rm bound}>10^7\rm~M_\odot$ across 24 realizations of host halos with $M_{\rm vir}(z_{\rm lens})=10^{13}\rm~M_{\odot}$ for PJ011646 and SDSS J0946+1006, whereas for JVAS B1938+666 we consider subhalos with bound masses $m_{\rm bound}>10^6\rm~M_\odot$ across 10 realizations. The right panels compare the posterior median density profile with those of a subhalo from each model family marked in the corresponding left panels. For PJ011646 and SDSS J0946+1006, we plot the density profiles of massive subhalos with enclosed masses consistent with the posteriors, while for JVAS B1938+666 we plot three subhalos with approximately the same enclosed mass and density slope. The shaded vertical gray bands highlight the range of three-dimensional radii over which spherical-shell density contributions account for the central $90\%$ of $M_{\rm 2D}(<R_{\rm pivot})$ for the median posterior density profiles. We note that the bands extend to radii beyond $R_{\rm pivot}$, since mass along the line of sight at three-dimensional radii $r>R_{\rm pivot}$ contributes to the projected enclosed mass $M_{\rm 2D}(<R_{\rm pivot})$.

\subsection{PJ011646}
\label{sec:PJ011646}

Under the fiducial SIDM model parameters, cored SIDM subhalos frequently reproduce the inferred structure of the perturber in PJ011646, whereas CDM subhalos only rarely do so and collapsed SIDM subhalos do not. As discussed in Section~\ref{sec:posteriors}, the posterior has a median projected enclosed mass $M_{\rm 2D}(<1.92\rm~kpc)\simeq3.7\times 10^9~\rm M_{\odot}$ and a median deflection-based projected density slope $\Gamma_{\rm 2D}(1.92\rm~kpc)\simeq-0.9$, as shown in panel~(a) of Figure~\ref{fig:pop_den}. To obtain more reliable counts of CDM and SIDM subhalos that lie within the approximate posterior region of the perturber, we generate 900 subhalo realizations under the fiducial model parameters $\sigma_0/m_{\chi}=147.1\rm~cm^2/g$ and $w=120\rm~km/s$, independent of the 24 realizations shown in Figure~\ref{fig:pop_den}. Within the rectangular region $3.0\times10^9\rm~M_\odot<M_{\rm 2D}(<1.92\rm~kpc)<5.0\times10^9\rm~M_\odot$ and $-1.1<\Gamma_{\rm 2D}(1.92\rm~kpc)<-0.75$, which approximately characterizes the 95\% posterior region, we find 2 CDM, $\sim550$ cored SIDM, and no collapsed SIDM subhalos. No collapsed subhalos lie within this region because their slopes at this pivot radius are steeper than $\Gamma_{\rm 2D}\simeq-1.2$ (Section~\ref{sec:proj_lensing_quantities}). To investigate the range of slopes exhibited by these massive subhalos, without attempting to represent the posterior region, we then widen the slope selection while keeping the mass selection fixed. For $-1.2<\Gamma_{\rm 2D}(1.92\rm~kpc)<-0.65$, we find $\sim100$ CDM, $\sim800$ cored SIDM, and still no collapsed SIDM subhalos. Further widening the selection to $-1.3<\Gamma_{\rm 2D}(1.92\rm~kpc)<-0.55$, which extends beyond the shallowest slopes of collapsed subhalos, yields $\sim600$ CDM, $\sim1000$ cored SIDM, and $\sim600$ collapsed SIDM subhalos.

The rarity of CDM candidates in our realizations is consistent with the high concentration inferred for this perturber. As the counts above show, massive CDM subhalos typically have slopes slightly shallower than the posterior. \cite{PJ011646_2026_dir_det_original} found $c_{200}=30^{+5}_{-7}$, about $3\sigma$ above the mass--concentration relation of CDM field halos~\citep{ludlow_2016_wdm}, and argued that this is consistent with a tidally stripped CDM subhalo based on the typical increase in concentration by a factor of $\approx1.5$ found for stripped subhalos in~\cite{springel_2008_aquarius}. For a subhalo of this mass, this factor implies $c_{200}\approx13$, still about a factor of two below the inferred value. Our CDM subhalos, whose tidal evolution is modeled with the transfer function of~\cite{Green_2019_tidal_tracks}, similarly only rarely reach the steep slope required at the pivot radius. Baryonic contraction, which our dark-matter-only realizations neglect, could steepen the inner profiles of CDM subhalos of this mass and increase the number of CDM candidates (Section~\ref{sec:discussion}).

Panel (b) compares the median posterior NFW density profile with those of the three subhalos marked by circles in panel (a). These subhalos have $M_{\rm 2D}(<1.92\rm~kpc)$ consistent with the posterior but different $\Gamma_{\rm 2D}(1.92\rm~kpc)$. The pivot radius therefore probes a qualitatively different region of the density profile for each model family. For the CDM subhalo, it lies within the scale radius of the profile, whereas for the cored and collapsed SIDM subhalos, it lies beyond their cored regions, where the density profiles are steeper. The shaded gray band illustrates why agreement in the enclosed mass within the pivot radius does not require agreement in the density profiles at smaller radii. The radial range over which density contributes to the central $90\%$ of $M_{\rm 2D}(<1.92\rm~kpc)$ for the posterior median lies largely outside the region where the selected profiles differ from the posterior.

\subsection{SDSS J0946+1006}
\label{sec:J0946}

Unlike for PJ011646, where a few CDM subhalos lie within the posterior region, we find no CDM subhalos within the posterior region of the perturber in SDSS J0946+1006, which simultaneously requires a large projected enclosed mass and a steep deflection-based projected density slope at a pivot radius of 0.63~kpc. However, collapsed SIDM subhalos frequently lie within the posterior region under the fiducial SIDM model parameters $\sigma_0/m_{\chi}=147.1\rm~cm^2/g$ and $w=120\rm~km/s$. Since these parameters were chosen in~\cite{Nadler_2023_sidm_sims} in part to produce core-collapsed subhalos relevant for this perturber (Section~\ref{sec:satgen_real_in_sidm}), we examine other parameter choices in Section~\ref{sec:vary_sidm_param}. As discussed in Section~\ref{sec:posteriors}, the posterior has a median $M_{\rm 2D}(<0.63\rm~kpc)\simeq2.2\times 10^9\rm~M_{\odot}$ and a median $\Gamma_{\rm 2D}(0.63\rm~kpc)\simeq-1.5$, as shown in panel (c) of Figure~\ref{fig:pop_den}. To obtain more reliable counts of CDM and SIDM subhalos that lie within the approximate posterior region of the perturber, we generate 900 subhalo realizations, independent of the 24 realizations shown in Figure~\ref{fig:pop_den}. Within the rectangular region $1.6\times10^9\rm~M_\odot<M_{\rm 2D}(<0.63\rm~kpc)<3.0\times10^9\rm~M_\odot$ and $-1.8<\Gamma_{\rm 2D}(0.63\rm~kpc)<-1.1$, which approximately characterizes the 95\% posterior region, we find no CDM, 4 cored SIDM, and $\sim2{,}600$ collapsed SIDM subhalos. To investigate the range of slopes exhibited by these massive subhalos, without attempting to represent the posterior region, we then widen the slope selection while keeping the mass selection fixed. For $-1.8<\Gamma_{\rm 2D}(0.63\rm~kpc)<-0.4$, we find no CDM, 42 cored SIDM, and the same $\sim2{,}600$ collapsed SIDM subhalos. CDM subhalos appear only once the shallow edge of the slope selection exceeds $\Gamma_{\rm 2D}(0.63\rm~kpc)\simeq-0.4$; for $-1.8<\Gamma_{\rm 2D}(0.63\rm~kpc)<-0.2$, we find $\sim70$ CDM, $\sim90$ cored SIDM, and again the same number of collapsed SIDM subhalos.

A CDM interpretation of the inferred perturber would require $\Gamma_{\rm 2D}(0.63\rm~kpc)\lesssim-1.1$, whereas all CDM subhalos of comparable enclosed mass in our realizations have $\Gamma_{\rm 2D}(0.63\rm~kpc)\gtrsim-0.4$, because the pivot radius lies well within their scale radii (Section~\ref{sec:proj_lensing_quantities}). Panel~(d) of Figure~\ref{fig:pop_den} illustrates the steep density slope required by the inferred perturber by comparing the posterior-median truncated NFW profile with the density profiles of three subhalos marked by squares in the corresponding left panel. The CDM and cored SIDM examples have enclosed masses consistent with the posterior but slopes that are too shallow at the pivot radius, whereas the collapsed SIDM example has both an enclosed mass and a slope consistent with the posterior. Because each SIDM subhalo is assumed to share the accretion and tidal evolution history of its CDM counterpart, the difference in compactness arises primarily from self-interactions: gravothermal collapse makes these subhalos more compact, while their CDM counterparts remain far less concentrated. 

\begin{figure}[t]
    \centering
    \includegraphics{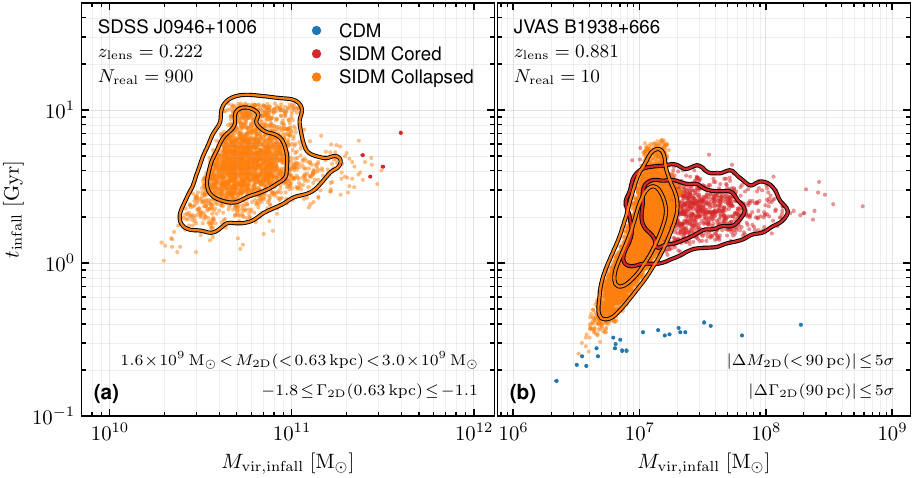}
    \caption{
    Infall properties of selected CDM and SIDM subhalos whose projected enclosed masses $M_{\rm 2D}(<R_{\rm pivot})$ (Eq.~\eqref{eq:M2D_integral_weighted}) and deflection-based projected density slopes $\Gamma_{\rm 2D}(R_{\rm pivot})$ (Eq.~\eqref{eq:gamma2d_deflection}) lie approximately within the posterior regions of the inferred perturbers in SDSS J0946+1006 and JVAS B1938+666, as shown in panels (c) and (e) of Figure~\ref{fig:pop_den}. Blue, red, and orange points denote CDM, cored SIDM, and collapsed SIDM subhalos, respectively, under the fiducial SIDM model parameters $\sigma_0/m_{\chi}=147.1\rm~cm^2/g$ and $w=120\rm~km/s$. \textbf{Panel (a):} infall time versus infall virial mass of subhalos with $1.6\times10^9\rm~M_{\odot}<M_{\rm 2D}(<0.63\rm~kpc)<3.0\times10^9\rm~M_{\odot}$ and $-1.8\leq\Gamma_{\rm 2D}(0.63\rm~kpc)\leq-1.1$ (SDSS J0946+1006; $z_{\rm lens}=0.222$, $R_{\rm pivot}=0.63$~kpc, $N_{\rm real}=900$ realizations). (Note that no CDM subhalos lie within the posterior of the perturber, and hence there are no blue points in this panel.) \textbf{Panel (b):} infall time versus infall virial mass of subhalos with $M_{\rm 2D}(<90\rm~pc)$ and $\Gamma_{\rm 2D}(90\rm~pc)$ lying within the $5\sigma$ posterior region of ``detection $\mathcal{V}$'' (JVAS B1938+666; $z_{\rm lens}=0.881$, $R_{\rm pivot}=90$~pc, $N_{\rm real}=10$ realizations). Contours enclose $68\%$ and $95\%$ of the selected SIDM subhalo populations.
    }
    \label{fig:progen_infall_plot}
\end{figure}

We further investigate the evolutionary histories of the SIDM subhalos that lie within the approximate 95\% posterior region of the inferred perturber across 900 realizations. Panel~(a) of Figure~\ref{fig:progen_infall_plot} shows the infall times and infall virial masses of the 4 cored and $\sim2{,}600$ collapsed subhalos. The 4 cored subhalos have core radii of $\simeq70\rm~pc$ and $\tau(z_{\rm lens})\simeq1.05$, just below the adopted collapse threshold $\tau(z_{\rm lens})=1.1$. Consequently, the pivot radius of 0.63~kpc lies well outside their core radii, where their density profiles are steep. They also have bound mass fractions $f_b\simeq0.15$, corresponding to a loss of about $85\%$ of their infall masses. The collapsed subhalos have a median infall virial mass $M_{\rm vir,infall}\simeq5.6\times10^{10}\rm~M_{\odot}$, a median infall time $t_{\rm infall}\simeq4.4\rm~Gyr$ (corresponding to $z_{\rm infall}\simeq1.4$) and a median bound mass fraction $f_b\simeq0.3$, corresponding to a loss of about $70\%$ of their infall masses. Relative to all SIDM subhalos (both cored and collapsed) within the enclosed mass selection, regardless of slope, these collapsed subhalos have median properties corresponding to the earliest $\sim50\%$ of infall times and lowest $\sim49\%$ of bound mass fractions. 
Our results suggest that the inferred structure of the perturber in SDSS J0946+1006 can frequently arise through the late-stage gravothermal evolution of massive subhalos, with infall masses of $\sim5.6\times10^{10}\rm~M_\odot$, without requiring unusual infall times or tidal histories.
At such infall masses, however, these subhalos would likely host a stellar component, which is included in neither the adopted lens model nor our dark-matter-only realizations (see the discussion of baryonic effects in Section~\ref{sec:discussion}). 

\subsection{JVAS B1938+666}
\label{sec:JVASB_analogs}

Under the fiducial SIDM model parameters $\sigma_0/m_{\chi}=147.1\rm~cm^2/g$ and $w=120\rm~km/s$, we find that SIDM subhalos frequently reproduce the inferred structure of ``detection $\mathcal{V}$'' in JVAS B1938+666, whereas CDM subhalos rarely do so and require unusual evolutionary histories. As discussed in Section~\ref{sec:posteriors}, the posterior of ``detection $\mathcal{V}$'' has a mean projected enclosed mass of $M_{\rm 2D}(<90~{\rm pc})\simeq1.2\times10^6~\rm M_{\odot}$ and a mean deflection-based projected density slope of $\Gamma_{\rm 2D}(90\rm~pc)\simeq-1.3$, as shown in panel~(e) of Figure~\ref{fig:pop_den}. Across 10 realizations, only 2 CDM, $\sim200$ cored SIDM, and $\sim2{,}200$ collapsed SIDM subhalos lie within the $2\sigma$ posterior regions in both $M_{\rm 2D}(<90~{\rm pc})$ and $\Gamma_{\rm 2D}(90\rm~pc)$. Within the $5\sigma$ posterior regions, we find 23 CDM, $\sim1{,}250$ cored SIDM, and $\sim10{,}200$ collapsed SIDM subhalos.
 
To examine which evolutionary histories produce these perturber candidates, we show in panel~(b) of Figure~\ref{fig:progen_infall_plot} the infall times and infall virial masses of the subhalos within the $5\sigma$ posterior regions. The collapsed subhalos have a median infall virial mass of $M_{\rm vir,infall}\simeq1.0\times10^7~\rm M_{\odot}$, a median infall time of $t_{\rm infall}\simeq1.5~\mathrm{Gyr}$ (corresponding to $z_{\rm infall}\simeq4$), and a median bound mass fraction of $f_b\simeq0.4$, corresponding to a loss of $\sim60\%$ of their infall masses. The cored subhalos have a median infall virial mass of $M_{\rm vir,infall}\simeq2.2\times10^7~\rm M_{\odot}$, a median infall time of $t_{\rm infall}\simeq2.2~\mathrm{Gyr}$ ($z_{\rm infall}\simeq3$), and a median bound mass fraction of $f_b\simeq0.3$, corresponding to a loss of $\sim70\%$ of their infall masses. They also have a median core radius of $\simeq10~\rm pc$ and a median $\tau(z_{\rm lens})\simeq1.0$, just below the adopted collapse threshold, so the 90~pc pivot radius lies well outside their cores, where $\Gamma_{\rm 2D}(90\rm~pc)$ is steep. The 23 CDM subhalos have a median infall time of $t_{\rm infall}\sim330$~Myr ($z_{\rm infall}\sim13$), a median infall virial mass of $M_{\rm vir,infall}\simeq7.8\times 10^6~\rm M_{\odot}$, and a median bound mass fraction of $f_b\simeq0.25$, corresponding to a loss of $\simeq75\%$ of their infall masses. At infall, before tidal stripping, their density profiles have a median $\Gamma_{\rm 2D}(90\rm~pc)\simeq-0.9$ and a median $M_{\rm 2D}(<90\rm~pc)\simeq2.5\times10^6\rm~M_\odot$, about twice the posterior mean, so none of them would lie within the $5\sigma$ posterior region without tidal evolution.

To assess how unusual these evolutionary histories are, we compare each sample with its ``parent'' population: subhalos of the same family that lie within the $5\sigma$ posterior region of $M_{\rm 2D}(<90\rm~pc)$ alone, without any selection on $\Gamma_{\rm 2D}(90\rm~pc)$. Only $\sim0.075\%$ of the CDM, $\sim6.6\%$ of the cored, and $\sim47\%$ of the collapsed parent subhalos are steep enough to also lie within the $5\sigma$ posterior region of $\Gamma_{\rm 2D}(90\rm~pc)$. Relative to their parent populations, the steep CDM subhalos have median properties corresponding to the earliest $\sim0.1\%$ of infall times and the lowest $\sim27\%$ of bound mass fractions, compared with the earliest $\sim22\%$ and lowest $\sim10\%$ for cored subhalos, and the earliest $\sim62\%$ and lowest $\sim76\%$ for collapsed subhalos. A CDM interpretation of ``detection $\mathcal{V}$'' therefore requires exceptionally early accretion, while the amount of tidal stripping is fairly typical; tidal stripping is necessary, since none of these subhalos would lie within the posterior region without it. A cored SIDM interpretation requires somewhat earlier infall and somewhat stronger tidal mass loss than typical, whereas a collapsed SIDM interpretation requires neither.  Panel~(f) of Figure~\ref{fig:pop_den} compares the posterior median density profile with one subhalo from each model family that approximately matches the posterior mean $M_{\rm 2D}(<90\rm~pc)$ and $\Gamma_{\rm 2D}(90\rm~pc)$, marked by the overlapping triangles in panel~(e). The three profiles illustrate the compact and steep structure required for consistency with ``detection $\mathcal{V}$''.

\begin{figure}[t]
    \centering
    \includegraphics{
    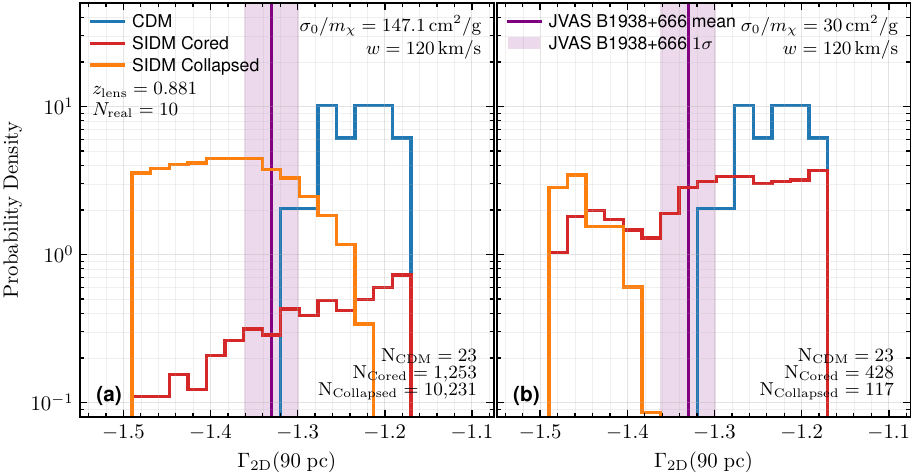}
    \caption{
    Probability density functions (PDFs) of the deflection-based projected density slope $\Gamma_{\rm 2D}(R_{\rm pivot})$ (Eq.~\eqref{eq:gamma2d_deflection}) for CDM and SIDM subhalos lying within the $M_{\rm 2D}(<90\rm~pc)$ and $\Gamma_{\rm 2D}(90\rm~pc)$ $5\sigma$ posterior regions of ``detection $\mathcal{V}$'' in JVAS B1938+666 ($z_{\rm lens}=0.881$, $R_{\rm pivot}=90$~pc, $N_{\rm real}=10$ realizations). The PDFs for cored and collapsed subhalos together integrate to unity, while the PDF for CDM subhalos alone integrates to unity. The numbers of selected CDM (blue), cored SIDM (red), and collapsed SIDM (orange) subhalos are indicated in the lower-right corner of each panel. \textbf{Panel (a):} PDFs of $\Gamma_{\rm 2D}(90\rm~pc)$ for selected subhalos under the fiducial SIDM model parameters $\sigma_0/m_{\chi}=147.1\rm~cm^2/g$ and $w=120\rm~km/s$, corresponding to the same subhalo samples in panel~(b) of Figure~\ref{fig:progen_infall_plot}. \textbf{Panel (b):} PDFs of $\Gamma_{\rm 2D}(90\rm~pc)$ for selected subhalos under $\sigma_0/m_{\chi}=30\rm~cm^2/g$ and $w=120\rm~km/s$. The solid purple vertical line and shaded band denote the posterior mean and its $1\sigma$ region, respectively. At the lower cross-section normalization, the SIDM subhalos closest to the posterior mean of ``detection $\mathcal{V}$'' are almost exclusively cored.
    }
    \label{fig:JVASB_gamma2d_dists}
\end{figure}

However, an SIDM interpretation of ``detection $\mathcal{V}$'' depends on the choice of SIDM model parameters $(\sigma_0/m_\chi,\,w)$. Panel~(a) of Figure~\ref{fig:JVASB_gamma2d_dists} shows the probability density functions (PDFs) of $\Gamma_{\rm 2D}(90\rm~pc)$ for the subhalos that lie within the $5\sigma$ posterior region under the fiducial SIDM model parameters, i.e., the same samples shown in panel~(b) of Figure~\ref{fig:progen_infall_plot}. In this case, collapsed subhalos dominate the SIDM sample near the posterior mean. Panel~(b) shows the corresponding PDFs when the cross-section normalization is reduced to $\sigma_0/m_{\chi}=30\rm~cm^2/g$ at fixed $w=120\rm~km/s$. Across the same 10 realizations, the number of SIDM subhalos within the $5\sigma$ posterior region then decreases by $\sim95\%$, from $\sim11{,}500$ to $\sim550$. At this lower cross-section normalization, the remaining collapsed subhalos all have $\Gamma_{\rm 2D}(90\rm~pc)\lesssim-1.37$, steeper than the $1\sigma$ region of the posterior, so the SIDM subhalos closest to the posterior mean are almost exclusively cored. Even so, SIDM subhalos still outnumber the 23 CDM subhalos by more than an order of magnitude, so the constraints on the SIDM model parameters in Section~\ref{sec:vary_sidm_param} are driven primarily by SDSS J0946+1006 rather than by ``detection $\mathcal{V}$''.

\subsection{Varying the SIDM model parameters}
\label{sec:vary_sidm_param}

\begin{figure}[p]
    \centering
    \includegraphics{
    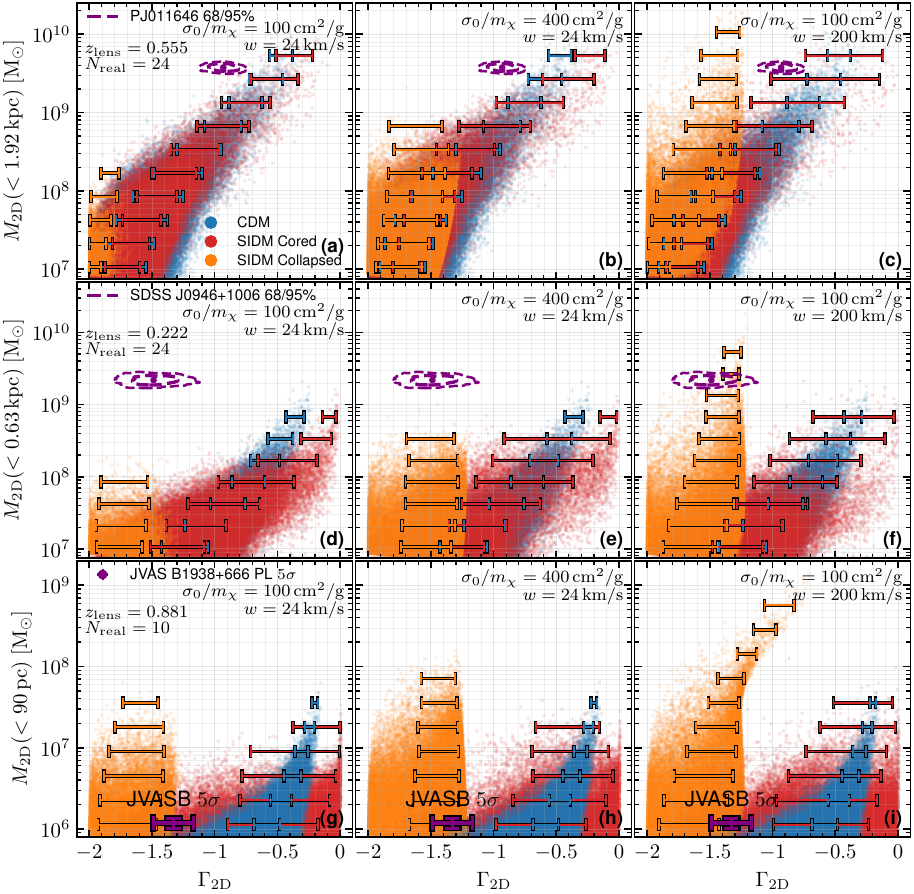}
    \caption{
    Comparison between the posteriors of inferred strong-lens perturbers and simulated CDM and SIDM subhalo populations, analogous to Figure~\ref{fig:pop_den}, with each column corresponding to a different set of SIDM model parameters. Blue, red, and orange points denote CDM, cored SIDM, and collapsed SIDM subhalos, respectively. \textbf{Top row:} PJ011646, $z_{\rm lens}=0.555$, $N_{\rm real}=24$, $R_{\rm pivot}=1.92$~kpc. \textbf{Middle row:} SDSS J0946+1006, $z_{\rm lens}=0.222$, $N_{\rm real}=24$, $R_{\rm pivot}=0.63$~kpc. \textbf{Bottom row:} JVAS B1938+666, $z_{\rm lens}=0.881$, $N_{\rm real}=10$, $R_{\rm pivot}=90$~pc. \textbf{Left column:} $\sigma_0/m_{\chi}=100\rm~cm^2/g$ and $w=24\rm~km/s$. \textbf{Middle column:} $\sigma_0/m_{\chi}=400\rm~cm^2/g$ and $w=24\rm~km/s$. \textbf{Right column:} $\sigma_0/m_{\chi}=100\rm~cm^2/g$ and $w=200\rm~km/s$. Horizontal bars denote the central $68\%$ interval of $\Gamma_{\rm 2D}$ in bins of $M_{\rm 2D}$ of width $0.3$~dex. Purple contours show the $68\%$ and $95\%$ posterior regions of the inferred perturbers in PJ011646 and SDSS J0946+1006; for ``detection $\mathcal{V}$'' in JVAS B1938+666, the purple marker shows the $5\sigma$ region. Among the parameter sets considered here and in Figure~\ref{fig:pop_den}, simultaneous overlap with all three posteriors requires a turnover velocity $w\gtrsim10^2\rm~km/s$, and is achieved for cross-section normalizations $\sigma_0/m_\chi\gtrsim10^2\rm~cm^2/g$.
    }
    \label{fig:pop_vary_sidm}
\end{figure}

We next investigate whether reproducing the structure of all three inferred perturbers with SIDM subhalos requires sufficiently large values of the SIDM model parameters. Figure~\ref{fig:pop_vary_sidm} repeats the comparison of Figure~\ref{fig:pop_den} for three alternative sets of parameters. Each row corresponds to one of the perturbers in PJ011646, SDSS J0946+1006, and JVAS B1938+666, while each column corresponds to one parameter set: $(\sigma_0/m_\chi,~w)=(100~{\rm cm^2/g},~24~{\rm km/s})$ (left), $(400~{\rm cm^2/g},~24~{\rm km/s})$ (middle), and $(100~{\rm cm^2/g},~200~{\rm km/s})$ (right). The CDM populations do not depend on these parameters and are therefore identical across columns. In the first column, cored SIDM subhalos lie within the PJ011646 posterior, no subhalos lie within the SDSS J0946+1006 posterior, and both cored and collapsed subhalos lie within the JVAS B1938+666 posterior. Reproducing the large enclosed mass of the SDSS J0946+1006 perturber requires massive progenitors (Section~\ref{sec:J0946}), whose internal velocities greatly exceed $w=24\rm~km/s$. Their self-interactions are therefore strongly velocity-suppressed, and they do not undergo sufficient gravothermal evolution to collapse. Lower-mass subhalos with smaller internal velocities still scatter efficiently and collapse, which is why the JVAS B1938+666 posterior remains populated. In the second column, quadrupling the cross-section normalization to $\sigma_0/m_\chi=400\rm~cm^2/g$ increases the abundance of collapsed subhalos, but massive subhalos still do not lie within the SDSS J0946+1006 posterior. For $v\gg w$, Eq.~\eqref{eq:sigma(v)} gives $\sigma_V\propto\sigma_0(w/v)^4$ up to a logarithmic factor, so a fourfold increase in $\sigma_0$ compensates for only a $\sim$40\% increase in velocity. In the third column, raising the turnover velocity to $w=200\rm~km/s$ at $\sigma_0/m_\chi=100\rm~cm^2/g$ greatly reduces the velocity suppression for massive subhalos, which can then collapse and populate the SDSS J0946+1006 posterior, while cored subhalos continue to lie within the PJ011646 posterior and both cored and collapsed subhalos lie within the JVAS B1938+666 posterior. Since the PJ011646 and JVAS B1938+666 posteriors are populated by SIDM subhalos for all parameter sets considered here, including $\sigma_0/m_\chi=30\rm~cm^2/g$ for JVAS B1938+666 (Section~\ref{sec:JVASB_analogs}), the requirement for simultaneous overlap is set by SDSS J0946+1006. Among the parameter sets considered here, including the fiducial set, populating the SDSS J0946+1006 posterior requires a turnover velocity of approximately $w\gtrsim10^2\rm~km/s$, and is achieved for cross-section normalizations $\sigma_0/m_\chi\gtrsim10^2\rm~cm^2/g$.

\subsection{Impact of alternative tidal stripping prescriptions in SIDM}
\label{sec:vary_sidm_strip}

\begin{figure}[p]
    \centering
    \includegraphics{
    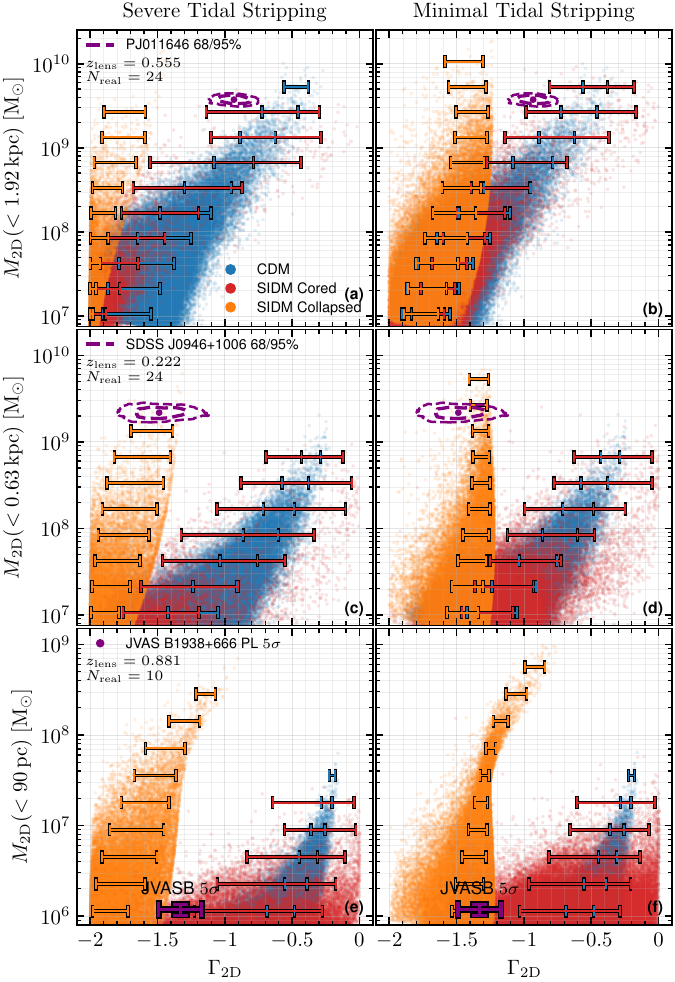}
    \caption{
    Comparison between the posteriors of inferred strong-lens perturbers and simulated CDM and SIDM subhalo populations, analogous to Figure~\ref{fig:pop_den}, for two alternative prescriptions for SIDM tidal stripping, under the fiducial SIDM parameters $\sigma_0/m_{\chi}=147.1\rm~cm^2/g$ and $w=120\rm~km/s$. \textbf{Left column:} severe tidal stripping, in which each SIDM subhalo retains only $10\%$ of the bound mass fraction of its CDM counterpart. \textbf{Right column:} minimal tidal stripping, in which the bound mass fraction is held fixed once an SIDM subhalo exceeds the collapse threshold $\tau(t)=1.1$. \textbf{Top row:} PJ011646, $z_{\rm lens}=0.555$, $N_{\rm real}=24$, $R_{\rm pivot}=1.92$~kpc. \textbf{Middle row:} SDSS J0946+1006, $z_{\rm lens}=0.222$, $N_{\rm real}=24$, $R_{\rm pivot}=0.63$~kpc. \textbf{Bottom row:} JVAS B1938+666, $z_{\rm lens}=0.881$, $N_{\rm real}=10$, $R_{\rm pivot}=90$~pc. Horizontal bars denote the central $68\%$ interval of $\Gamma_{\rm 2D}$ in bins of $M_{\rm 2D}$ of width $0.3$~dex. Our conclusions from Section~\ref{sec:results:perturbers} are robust to these alternative stripping prescriptions: SIDM subhalos reproduce the structures of the inferred perturbers more frequently than CDM subhalos under both prescriptions.
    }
    \label{fig:vary_sidm_strip}
\end{figure}

As discussed in Section~\ref{sec:results:perturbers} and shown in Figure~\ref{fig:pop_den}, under the fiducial SIDM model parameters $\sigma_0/m_{\chi}=147.1\rm~cm^2/g$ and $w=120\rm~km/s$, cored SIDM subhalos lie within the 95\% posterior region of the PJ011646 perturber and both cored and collapsed SIDM subhalos lie within the 95\% posterior region of the SDSS J0946+1006 perturber and the $5\sigma$ posterior region of ``detection $\mathcal{V}$'' in JVAS B1938+666. These results assume the fiducial prescription for tidal mass loss in SIDM adopted thus far, in which every SIDM subhalo retains the same fraction of its infall mass as its CDM counterpart. In Figure~\ref{fig:vary_sidm_strip}, we consider two alternative tidal stripping prescriptions for SIDM subhalos that bracket this assumption, while the tidal evolution of CDM subhalos remains unchanged. These prescriptions modify only the bound mass fractions, and hence the truncation radii, of SIDM subhalos, not their gravothermal evolution histories. Since stronger tidal stripping would also accelerate gravothermal evolution, the severe-stripping scenario may underestimate the number of collapsed subhalos.

The left column shows a ``severe'' tidal stripping scenario in which each SIDM subhalo retains only $10\%$ of the bound mass fraction of its CDM counterpart. This choice is motivated by the idealized SIDM $N$-body simulations of~\cite{klemmer_2026_sidm_survival}, which adopt the same velocity-dependent SIDM model used in our analysis~\cite{Yang_2022_viscosity_cs_sims}, but with $\sigma_0/m_\chi=90~\rm cm^2/g$ and $w=188~\mathrm{km/s}$. In their simulations, a $10^{10}~\mathrm{M_\odot}$ subhalo evolved for $5~\mathrm{Gyr}$ within a $10^{13}~\mathrm{M_\odot}$ host on an orbit with an apocenter of $2r_{s,\rm host}$ and a pericenter of $r_{s,\rm host}$ retained about an order of magnitude less bound mass than its CDM counterpart; see their Figure~4~\citep{klemmer_2026_sidm_survival}. As shown in the left column of Figure~\ref{fig:vary_sidm_strip}, severe stripping reduces the projected enclosed masses $M_{\rm 2D}(<R_{\rm pivot})$ and steepens the deflection-based projected density slopes $\Gamma_{\rm 2D}(R_{\rm pivot})$, because the truncation radii of many subhalos approach or fall below the pivot radius, driving their $\Gamma_{\rm 2D}(R_{\rm pivot})$ closer to the point-mass limit (Section~\ref{sec:proj_lensing_quantities}). The right column shows a ``minimal'' tidal stripping scenario in which the bound mass fraction of an SIDM subhalo is held fixed once its gravothermal evolution parameter exceeds the collapse threshold $\tau(t)=1.1$. This choice is motivated by cosmological SIDM zoom-in simulations showing that core-collapsed subhalos are more resilient to tidal disruption than cored ones~\cite{Nadler_2023_sidm_sims}, and by idealized $N$-body simulations showing that tidal mass loss slows as cuspy subhalos approach stable bound remnants~\cite{Errani_2022_survival}. We therefore treat this scenario as a limiting case of strongly suppressed tidal stripping in SIDM, in which mass loss stops entirely after collapse. With larger truncation radii, collapsed subhalos exhibit shallower $\Gamma_{\rm 2D}(R_{\rm pivot})$, as shown in the right column of Figure~\ref{fig:vary_sidm_strip}.

For each alternative prescription, we compare the number of subhalos lying approximately within the posterior regions of the three perturbers with that under the fiducial prescription. As in Sections~\ref{sec:PJ011646}--\ref{sec:JVASB_analogs}, we use the 900 realizations for PJ011646 and SDSS J0946+1006, which are independent of the 24 realizations shown in Figure~\ref{fig:vary_sidm_strip}, and the 10 realizations shown in the figure for JVAS B1938+666. Since minimal stripping applies only to subhalos with $\tau(t)>1.1$, the cored populations are also unchanged under this prescription.

Under severe stripping, the number of cored subhalos that lie within the PJ011646 posterior region decreases by $\sim90\%$, to $\approx65$, compared with 2 CDM subhalos. For SDSS J0946+1006, the number of collapsed subhalos decreases by $\sim70\%$, to $\approx700$, compared with no CDM subhalos, and the 4 cored subhalos discussed in Section~\ref{sec:J0946} no longer lie within the posterior region. For ``detection $\mathcal{V}$'' in JVAS B1938+666, the numbers of cored and collapsed subhalos decrease by $\sim55\%$ and $\sim99\%$, to $\approx560$ and $\approx65$, respectively, compared with 23 CDM subhalos. In this case, cored subhalos lie within the posterior region more frequently than collapsed subhalos, and have a median $\tau(z_{\rm lens})\simeq1.04$, just below the adopted collapse threshold. Under minimal stripping, the number of collapsed subhalos that lie within the posterior region increases by $\sim20\%$ for SDSS J0946+1006 and by $\sim35\%$ for JVAS B1938+666, and remains zero for PJ011646. Although severe stripping substantially reduces the number of SIDM perturber candidates, our conclusions from Section~\ref{sec:results:perturbers} are robust to both prescriptions: SIDM subhalos reproduce the inferred structures of all three perturbers more frequently than CDM subhalos.

\section{Discussion of modeling uncertainties}
\label{sec:discussion}

In this section, we discuss four modeling uncertainties that may affect our predictions and conclusions. Addressing these uncertainties in future work will allow for systematic improvements to our framework.

\textbf{Late-stage gravothermal evolution:} The detailed behavior of late-stage gravothermal evolution remains theoretically uncertain, and accurately modeling this regime is an active area of research. The gravothermal evolution model adopted in this work is calibrated only through the onset of core collapse and can underestimate the central density of halos by $\sim15\%$ at $\tau(t)=1.0$~\cite{Yang_2024_parametric}. This can result in underestimated projected enclosed masses and inaccurate deflection-based projected density slopes for subhalos with $1\lesssim\tau(t)\lesssim1.1$ that are assigned the cored profile in Eq.~\eqref{eq:daneng_prof}. For $\tau(t)>1.1$, we assign SIDM subhalos the gravothermally collapsed profile from~\cite{Slone_2026}, which has a three-dimensional inner power-law slope outside its core of $\gamma\sim2.2$. This allows us to better approximate the lensing-relevant structure of collapsed subhalos without extrapolating the gravothermal evolution model beyond its calibrated range. Therefore, our binary assignment of density profiles to cored and collapsed subhalos in this work should be interpreted as an approximation of different stages of gravothermal evolution, rather than as a prediction of two distinct populations. In addition, the kinetic simulations of gravothermal evolution in~\cite{gurian_may_deep_col} found an inner three-dimensional logarithmic density slope of $\gamma\sim2.6$ at a late stage of gravothermal collapse, corresponding to $\tau(t)\simeq12.2$ under the collapse timescale definition in Eq.~\eqref{eq:t_c_original}. In Appendix~\ref{app:oren_gurian_compare}, we demonstrate that our conclusions in Section~\ref{sec:results:perturbers} are robust when assigning this alternative collapsed profile to subhalos with $\tau(t)>1.1$. However, as discussed in the appendix and shown in Figure~\ref{fig:density_profiles}, the two collapsed profiles differ at radii $\lesssim0.01r_s$, affecting the distributions of projected enclosed masses and deflection-based projected density slopes of collapsed subhalos with larger masses and smaller pivot radii. Improved modeling of late-stage gravothermal evolution would yield more reliable predictions for the projected structures of SIDM subhalos and enable more robust comparisons with inferred strong-lens perturbers.

\textbf{Tidal evolution, mass loss, and survival in SIDM:} We assume that each SIDM subhalo retains the same fraction of its infall mass as its CDM counterpart by applying a time-dependent truncation radius described in Section~\ref{sec:subhalo_den_profiles}. This approach approximates the tidal evolution of SIDM subhalos while their internal density profiles evolve according to the gravothermal evolution model, but it does not capture how the evolving SIDM density profiles in turn affect their mass-loss histories. Two effects are therefore missing. First, the response of a subhalo to the host tidal field depends on its density profile: cored subhalos with shallower gravitational potentials are more vulnerable to tidal stripping and disruption~\cite{Errani_2022_survival,klemmer_2026_sidm_survival}, whereas core-collapsed subhalos are more resilient~\citep{Nadler_2023_sidm_sims}. Second, scattering between subhalo and host particles can cause additional mass loss and modify gravothermal evolution~\citep{Nadler_2020_sidm_mass_loss,klemmer_2026_sidm_survival,Zheng_2022_life_story}. This effect is strongest when the cross section remains large at the relative velocities between subhalo and host particles, as for the large turnover velocities favored in Section~\ref{sec:vary_sidm_param}. The importance of both effects depends nontrivially on the subhalo orbit and density profile, the relative velocities between subhalo and host particles, and the velocity and angular dependence of the self-interaction cross section~\citep{kummer_2017_sidm_mass_loss, slone_2023_sidm_param_draco, shirasaki_2022_sidm_evo}. In addition, we assign an SIDM counterpart to every surviving CDM subhalo, whereas self-interactions can alter which subhalos survive~\citep{Nadler_2020_sidm_mass_loss}. The ``severe'' and ``minimal'' tidal stripping scenarios of Section~\ref{sec:vary_sidm_strip} provide approximate bounds on these effects, but a self-consistent treatment of mass loss and survival in SIDM is left to future work.

\textbf{Baryonic effects in CDM and SIDM:} All subhalos in this work are evolved in dark-matter-only hosts, without accounting for the baryonic potential of the host galaxy or hydrodynamic processes such as ram-pressure stripping, which can affect subhalo evolution and survival~\citep{richings_2021_baryons_strong_lens_n_body,kimmel_2017_hydro_survival_fire,despali_2017_baryons_strong_lens}. Baryons can also change the internal structure of subhalos. Hydrodynamic cosmological simulations show that baryonic processes can increase the central densities and concentrations of CDM field halos and subhalos and steepen their inner density slopes relative to dark-matter-only predictions~\citep{duffy_2010_hydro_profiles,zhu_2016_baryonic_impact,heinz_tng50_baryonic_contract,gnedin_2004_bary_contract}. In particular, \cite{heinz_tng50_baryonic_contract} found that most CDM subhalos with masses $\gtrsim10^8\rm~M_\odot$ in the TNG50 simulation have three-dimensional logarithmic inner density slopes of $\gamma\sim2$. Their dark matter components exhibit similarly steep slopes due to their gravitational response to baryons. Such steepening would shift massive CDM subhalos toward steeper $\Gamma_{\rm 2D}$ values than in our dark-matter-only realizations, which could make CDM interpretations of steep perturbers less rare. This is particularly relevant for the PJ011646 perturber, whose slope is only slightly steeper than those of typical CDM subhalos of comparable mass (Section~\ref{sec:PJ011646}). However, \cite{Minor_2021_J0946_concentration} found that the TNG100 subhalos steep enough to match the inferred perturber in SDSS J0946+1006 owe their slopes to adiabatic contraction by stellar masses an order of magnitude above the upper limit implied by the perturber's inferred luminosity. In SIDM halos, baryons likewise modify the central dark matter profile, with a response that depends on the baryonic mass distribution and the strength of self-interactions~\citep{despali_2019_sidm_baryons,sameie_2021_sidm_baryons}.

Baryons also affect the lensing inference itself, since stellar mass and light associated with a perturber can alter its inferred projected enclosed mass and density slope~\citep{He_2025_not_so_dark,Minor_2021_J0946_concentration}. The collapsed SIDM subhalos that lie approximately within the SDSS J0946+1006 posterior have a median infall mass of $\simeq5.6\times10^{10}\rm~M_\odot$ (Section~\ref{sec:J0946}), for which a stellar population is plausible~\citep{Li_2025_dark_dot,Ballard_2024_J0946,Tajalli_2025_J0946_sidm}. Such a population is included in neither the gravitational-imaging analysis adopted in this paper~\citep{minor_2025_direct_detect} nor our subhalo realizations, whereas \cite{He_2025_not_so_dark} explicitly modeled a luminous perturber and showed that neglecting its light can substantially bias the inferred structure. Baryonic effects are therefore most important for the massive subhalos relevant to PJ011646 and SDSS J0946+1006. In contrast, the CDM and SIDM subhalos consistent with ``detection $\mathcal{V}$'' in JVAS B1938+666 have median infall masses of $\sim10^7\rm~M_\odot$ (Section~\ref{sec:JVASB_analogs}), well below the mass scale at which star formation becomes efficient~\citep{gao_2004_subhalos,sawala_2016_low_subhalos}, so their own baryonic content is likely negligible, although the baryonic potential of the host can still affect their tidal evolution.

\textbf{Host halo mass of strong-lens systems:} The abundance of subhalos depends on the assumed host halo mass~\citep{han_2016_subhalo_mass_func}, which may differ among the strong-lens systems considered in this work. For all three systems, we assume a host virial mass of $M_{\rm vir}(z_{\rm lens})=10^{13}~\rm M_\odot$, representative of the $\sim10^{13}~\rm M_\odot$ group-scale halos associated with the massive early-type galaxies that dominate galaxy-scale strong-lens samples~\citep{Vegetti_2023_strong_lens_dm_review,SL_host_mass_measure_slacs}. Strong-lensing observations primarily constrain the mass distribution within the Einstein radius, while the virial mass of the host halo is typically inferred indirectly and remains uncertain~\citep{enzi_2021_wdm_strong_lens}. For SDSS J0946+1006, \cite{Li_J0946_host_mass} combined lensing, stellar kinematics, and a concentration--mass prior to infer $M_{200}=1.11^{+0.37}_{-0.32}\times10^{13}~\rm M_\odot$, although this measurement is model-dependent and requires extrapolation beyond the radial range directly constrained by lensing. In~\cite{Gilman_2019_wdm_chills}, the host-mass and redshift dependence of the projected subhalo mass function normalization was parametrized through the scaling function
\begin{equation}
\log_{10}\mathcal{F}=0.88\log_{10}\left(
\frac{M_{\rm host}}{10^{13}\mathrm{M}_\odot}\right)
+1.7\log_{10}(z_{\rm lens}+0.5),
\end{equation}
which is calibrated against the semi-analytic model \textsc{Galacticus}~\citep{Benson_2012_galacticus}. At fixed redshift, the projected subhalo abundance scales approximately linearly with host halo mass, $\mathcal{F}\propto M_{\rm host}^{0.88}$. Therefore, the predicted number of subhalos that lie within the posteriors of the three perturbers is sensitive to the assumed host halo mass of each lens system. However, because each SIDM subhalo is mapped from a CDM counterpart, a different host halo mass rescales the abundances of CDM and SIDM subhalos together to first order, so the relative frequency of CDM and SIDM perturber candidates is less sensitive to this assumption than their absolute numbers.

\section{Conclusions}
\label{sec:conclusion}

We combined semi-analytic CDM subhalo histories with a model for gravothermal evolution in SIDM to efficiently generate CDM and SIDM subhalo populations across SIDM parameter space and evaluate their projected enclosed masses and deflection-based projected density slopes. We compared these populations with three perturbers inferred through gravitational imaging in PJ011646, SDSS J0946+1006, and JVAS B1938+666, at the pivot radii around which their inferred structures are most robustly constrained. We found that SIDM can produce structures consistent with the three inferred perturbers more frequently than CDM. Under the fiducial SIDM model parameters $\sigma_0/m_{\chi}=147.1\rm~cm^2/g$ and $w=120\rm~km/s$, a strong-lens host contains on average $\sim0.002$ CDM, $\sim0.6$ cored SIDM, and no collapsed SIDM subhalos that reproduce the inferred structure of the PJ011646 perturber at a pivot radius of $\simeq1.92\rm~kpc$ (based on 900 realizations), since collapsed subhalos of comparable mass have steeper slopes than the perturber. In contrast, for SDSS J0946+1006, we find no CDM subhalos in 900 realizations, compared with $\sim0.004$ cored and $\sim3$ collapsed SIDM subhalos per host that reproduce the large enclosed mass and steep slope of the perturber at a pivot radius of $\simeq0.63\rm~kpc$. For ``detection $\mathcal{V}$'' in JVAS B1938+666, a host contains on average $\sim0.2$ CDM, $\sim20$ cored SIDM, and $\sim220$ collapsed SIDM subhalos within the $2\sigma$ posterior region at a pivot radius of $\simeq90\rm~pc$ (based on 10 realizations). These per-host numbers refer to subhalos anywhere within the host virial radius; the number projected near the lensed images is much smaller (Appendix~\ref{app:ann_num_den_appendix}). Within the wider $5\sigma$ region, the 23 CDM subhalos found across all realizations require unusually early accretion: relative to all CDM subhalos with projected enclosed masses within $5\sigma$ of ``detection $\mathcal{V}$'', their median infall time corresponds to the earliest $\sim0.1\%$, whereas their bound mass fractions are less unusual (lowest $\sim27\%$). When varying the SIDM model parameters, we found that simultaneously reproducing the inferred structure of all three perturbers requires a turnover velocity of approximately $w\gtrsim10^2\rm~km/s$, so that the massive progenitors of SDSS J0946+1006 perturber candidates can gravothermally collapse, and is achieved for cross-section normalizations $\sigma_0/m_\chi\gtrsim10^2\rm~cm^2/g$ among the parameter sets considered. Our results are robust to the alternative tidal stripping prescriptions considered in Section~\ref{sec:vary_sidm_strip}, although the abundance of SIDM subhalos that reproduce the inferred structure of the perturbers is substantially suppressed under severe tidal stripping. 

These conclusions are subject to the modeling uncertainties discussed in Section~\ref{sec:discussion}. In particular, the collapsed SIDM subhalos that reproduce the SDSS J0946+1006 perturber have a median infall mass of $\simeq5.6\times10^{10}\rm~M_\odot$, for which a stellar component is plausible but is included in neither the adopted lens model nor our dark-matter-only realizations. Baryonic contraction, also neglected here, could steepen the inner profiles of massive CDM subhalos and increase the number of CDM candidates for the more massive perturbers. In addition, the abundance of SIDM perturber candidates depends on the treatment of late-stage gravothermal evolution and of tidal evolution in SIDM (Section~\ref{sec:vary_sidm_strip}). The framework developed here can be systematically improved by addressing these effects. For individual perturber detections, such improvements will sharpen comparisons between a detected object and the population of subhalos from which it is drawn, quantifying how typical its inferred structure is under a given dark matter model. For flux-ratio anomalies and other population-level inferences, they will enable more robust predictions of the lensing signal expected for a given SIDM model. 

Both individual perturber detections and population-level inferences will become increasingly powerful as current and upcoming surveys substantially increase the known strong-lens population. The Vera C. Rubin Observatory is forecast to discover tens of thousands of galaxy--galaxy lenses, while the Euclid mission is expected to identify $\sim10^5$ galaxy-scale strong lenses over its full survey~\citep{Collett_2015_rubin_strong_lens_forecast, Euclid_2026_Q1, shajib_2025_rubin}. Forecasts for the James Webb Space Telescope suggest hundreds of lenses per square degree in sufficiently deep observations~\citep{silver_2025_jwst_strong_lens_forecast}. The Nancy Grace Roman Space Telescope is predicted to discover $\sim160{,}000$ strong lenses, of which $\sim500$ may have sufficient signal-to-noise for substructure inference through gravitational imaging~\citep{Wedig_2025_Roman_forecasts}, with further deep-field observations in the extended lifetime of the telescope substantially increasing the expected sample. Together, these facilities will greatly increase the number of lens systems in which dark matter perturbers can potentially be discovered. Because our framework can efficiently predict the distributions of subhalo projected enclosed masses and deflection-based projected density slopes across SIDM parameter space, lens redshifts, and pivot radii, it is well suited for statistical comparisons between future perturber samples and the predictions of different dark matter models. 

\acknowledgments

We thank Oren Slone for useful discussions and for providing the shape of the collapsed profile. RE acknowledges support from the DOE Grant DE-SC0025309 and the Simons Investigator in Physics award MPS-SIP-00010469. DA and JT were supported in part by DOE Grant DE-SC0025309 and the Simons Investigator in Physics award MPS-SIP-00010469. DA is also supported by NSF 2514888. JT and SB also acknowledge support from DOE Grant DE-SC0026113. DG gratefully thanks the Brinson Foundation for their continued support of his research, and acknowledges support from the Kavli Foundation under Grant No. PS-2025-GR-249-3072. ChatGPT (OpenAI) and Claude (Anthropic) were occasionally used in developing and testing code, making suggestions for clarity, grammar, and spelling, and checking the manuscript and citations for errors; the authors verified all output and take full responsibility for the content of this work. 

\appendix

\section{Annular number density of subhalos with perturber-like enclosed masses}
\label{app:ann_num_den_appendix}

\begin{figure}[t]
    \centering
    \includegraphics{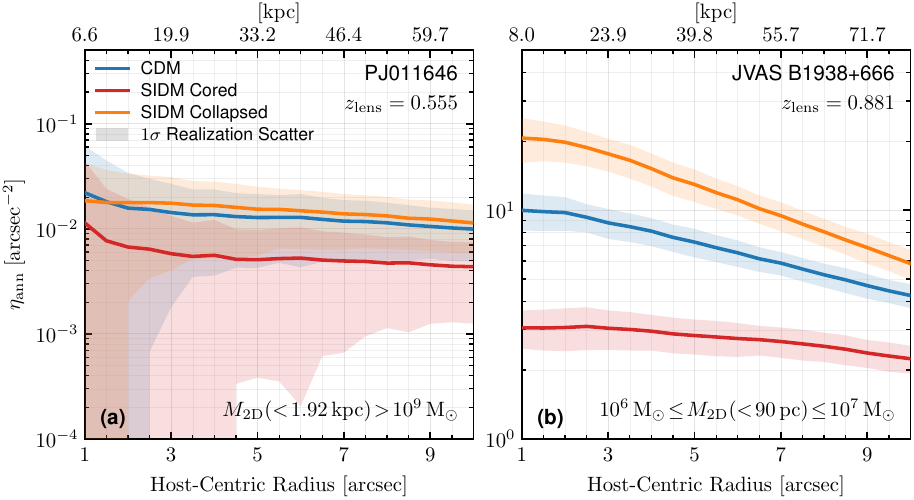}
    \caption{
    Orientation-averaged annular number densities (Eq.~\eqref{eq:annular_number_density}) of mass-selected CDM and SIDM subhalos versus projected host-centric radius for PJ011646 and JVAS B1938+666, under the fiducial SIDM model parameters $\sigma_0/m_{\chi}=147.1\rm~cm^2/g$ and $w=120\rm~km/s$. Subhalos are selected by their projected enclosed mass within the pivot radius alone, with no selection on $\Gamma_{\rm 2D}(R_{\rm pivot})$, so these densities also include subhalos whose slopes lie outside the posterior regions of the inferred perturbers. Blue, red, and orange curves denote the annular number densities of CDM, cored SIDM, and collapsed SIDM subhalos, respectively. \textbf{Panel (a):} PJ011646 ($z_{\rm lens}=0.555$, $R_{\rm pivot}=1.92$~kpc), selecting subhalos with $M_{\rm 2D}(<1.92\rm~kpc)>10^9\rm~M_{\odot}$, averaged over $N_{\rm real}=900$ realizations. \textbf{Panel (b):} JVAS B1938+666 ($z_{\rm lens}=0.881$, $R_{\rm pivot}=90$~pc), selecting subhalos with $10^6\rm~M_{\odot}\leq M_{\rm 2D}(<90\rm~pc)\leq10^7\rm~M_{\odot}$, averaged over $N_{\rm real}=10$ realizations. The annular width is fixed at $0.5$~arcsec, with the host-centric radius corresponding to the midpoint of an annulus. Shaded bands denote the $1\sigma$ realization scatter. The upper axes show the corresponding host-centric radius in projected physical distances at the lens redshift of each system.
    }
    \label{fig:num_den_plot}
\end{figure}

In this appendix, we evaluate the orientation-averaged annular number densities of mass-selected CDM and SIDM subhalos relevant to the inferred perturbers in PJ011646 and JVAS B1938+666, under the fiducial SIDM model parameters $\sigma_0/m_{\chi}=147.1\rm~cm^2/g$ and $w=120\rm ~km/s$. For a subhalo at three-dimensional host-centric radius $r_i$, the probability that a random viewing orientation places it within a projected annulus $R_{\rm in}<R<R_{\rm out}$ is
\begin{equation}
P_{i}=W(R_{\rm out}/r_i)-W(R_{\rm in}/r_i),
\label{eq:annular_projection_probability}
\end{equation}
where $W$ is the geometric projection kernel defined in Eq.~\eqref{eq:kernel-cyl-sphere}. The number density within the annulus is then
\begin{equation}
\eta_{\rm ann}=\frac{\displaystyle\sum_{i}P_i}
{\pi\left(R_{\rm out}^{2}-R_{\rm in}^{2}\right)}.
\label{eq:annular_number_density}
\end{equation}
In Figure~\ref{fig:num_den_plot}, we show the annular number densities of mass-selected CDM and SIDM subhalos using a fixed annular width of $0.5~\rm arcsec$, with each plotted host-centric radius representing the midpoint of an annulus. Subhalos are selected by their projected enclosed mass within the pivot radius of each inferred perturber (Section~\ref{sec:posteriors}), with no selection based on their deflection-based projected density slope $\Gamma_{\rm 2D}$ (Section~\ref{sec:proj_lensing_quantities}) because the enclosed mass of the inferred perturbers is more robustly constrained than their projected density slopes~\citep{PJ011646_2026_dir_det_original,Vegetti_2026_cdm_unexpect}. Therefore, these annular number densities predict the projected abundances of CDM and SIDM subhalos capable of supplying the relevant enclosed masses. These predicted abundances should not be interpreted as detection rates, which depend on the subhalo positions relative to lensed images and the lensing selection functions determined by data quality and the adopted lens and source models~\citep{Vegetti_2023_strong_lens_dm_review, Nightingale_2024_subhalo_scan}.

Panel (a) selects subhalos with projected enclosed masses $M_{\rm 2D}(<1.92~\rm kpc)>10^9~\rm M_\odot$ at $z_{\rm lens}=0.555$ relevant for the inferred perturber in PJ011646. The fraction of SIDM subhalos under this mass selection that are collapsed at the lens redshift is $\sim62\%$, while the remainder are cored. Averaged over 900 realizations, the annular number densities of mass-selected CDM and collapsed SIDM subhalos are comparable, $\sim1$–$2\times10^{-2}\rm~arcsec^{-2}$, while those of cored SIDM subhalos are lower by a factor of $\sim2$, demonstrating the rarity of these massive subhalos within the annulus. The realization-scatter is comparable to or larger than the difference between the model families. These annular densities reflect the mass selection alone. When the slope is also required to lie within the posterior region (Section~\ref{sec:PJ011646}), the ranking of the three families changes: collapsed subhalos are too steep and CDM subhalos are mostly too shallow, whereas cored subhalos, despite being the least abundant in this mass-selected sample, most often have slopes consistent with the inferred perturber.

On the other hand, for ``detection $\mathcal{V}$'' in JVAS B1938+666, the differences between the annular number densities of CDM, cored, and collapsed subhalos exceed the realization-scatter under the fiducial SIDM model parameters. Panel (b) selects subhalos with enclosed masses $10^{6}\rm~M_\odot<M_{\rm 2D}(<90~\rm pc)<10^{7}\rm~M_\odot$ at $z_{\rm lens}=0.881$. Under this mass selection, the fraction of SIDM subhalos that are collapsed at $z_{\rm lens}=0.881$ is $\sim66\%$, while the rest are cored. Across 10 realizations, the annular number densities at $\simeq1~\rm arcsec$ from the host center are $\sim10~\rm arcsec^{-2}$ for CDM subhalos, while their cored and collapsed counterparts have densities $\sim3~\rm arcsec^{-2}$ and $\sim20~\rm arcsec^{-2}$, respectively. The number densities in JVAS B1938+666 are orders of magnitude larger than those in PJ011646, primarily because the mass selection probes much lower mass, more abundant subhalos. In addition, the fixed annular width of $0.5$~arcsec corresponds to a larger projected physical distance at $z_{\rm lens}=0.881$ ($\simeq4$~kpc) than at $z_{\rm lens}=0.555$ ($\simeq3.3$~kpc). For the low-mass subhalos relevant for the inferred perturber in JVAS B1938+666 under the fiducial SIDM model parameters, the annular abundance of collapsed subhalos is significantly larger than that of CDM or cored subhalos.

\section{Comparison of collapsed SIDM profiles}
\label{app:oren_gurian_compare}

\begin{figure}[p]
    \centering
    \includegraphics{
    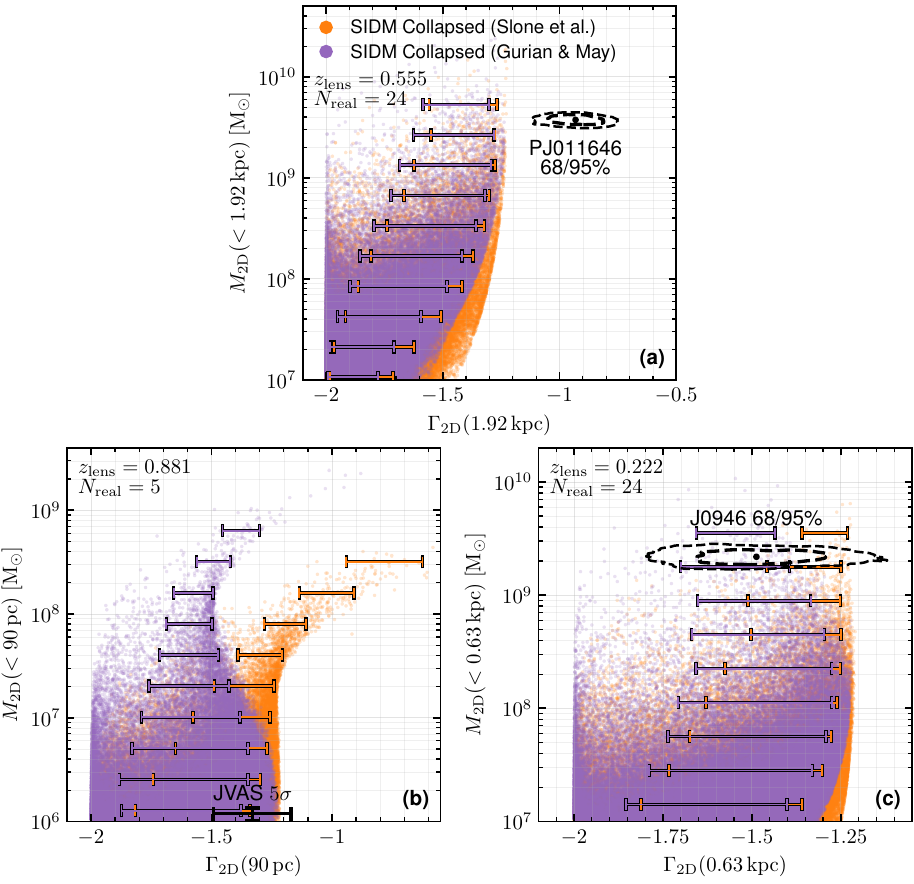}
    \caption{
    Comparison of the projected enclosed masses $M_{\rm 2D}(<R_{\rm pivot})$ (Eq.~\eqref{eq:M2D_integral_weighted}) and deflection-based projected density slopes $\Gamma_{\rm 2D}(R_{\rm pivot})$ (Eq.~\eqref{eq:gamma2d_deflection}) of collapsed SIDM subhalos and inferred strong-lens perturbers, using two different collapsed density profiles and the same fiducial tidal truncation prescription described in Section~\ref{sec:subhalo_den_profiles}, under the fiducial SIDM model parameters $\sigma_0/m_{\chi}=147.1\rm~cm^2/g$ and $w=120\rm~km/s$. Orange denotes subhalos assigned a collapsed density profile from~\cite{Slone_2026}, and purple denotes subhalos assigned a collapsed density profile from Figure 1 of~\cite{gurian_may_deep_col}. \textbf{Panel~(a):} $M_{\rm 2D}(<1.92\rm~kpc)$ versus $\Gamma_{\rm 2D}(1.92\rm~kpc)$ for $N_{\rm real}=24$ realizations alongside the posterior region of the inferred perturber in PJ011646~\cite{PJ011646_2026_dir_det_original} ($z_{\rm lens}=0.555$, $R_{\rm pivot}=1.92$~kpc). \textbf{Panel (b):} $M_{\rm 2D}(<90\rm~pc)$ versus $\Gamma_{\rm 2D}(90\rm~pc)$ for $N_{\rm real}=5$ realizations alongside the posterior (PL $5\sigma$ region) of detection $\mathcal{V}$ in JVAS B1938+666~\cite{Vegetti_2026_cdm_unexpect} ($z_{\rm lens}=0.881$, $R_{\rm pivot}=90$~pc). \textbf{Panel (c):} $M_{\rm 2D}(<0.63\rm~kpc)$ versus $\Gamma_{\rm 2D}(0.63\rm~kpc)$ for $N_{\rm real}=24$ realizations alongside the posterior region of the inferred perturber in SDSS J0946+1006~\cite{minor_2025_direct_detect} ($z_{\rm lens}=0.222$, $R_{\rm pivot}=0.63$~kpc). Horizontal bars denote the central $68\%$ interval of $\Gamma_{\rm 2D}$ in bins of $M_{\rm 2D}$ of width $0.3$~dex. Our conclusions from Section~\ref{sec:results:perturbers} are robust to the choice of collapsed density profile. 
    }
    \label{fig:oren_gurian_compare}
\end{figure}

In Figure~\ref{fig:oren_gurian_compare}, we test the sensitivity of the projected structure of SIDM subhalos to the choice of gravothermally collapsed density profile assigned to subhalos that have exceeded the adopted collapse threshold $\tau(z_{\rm lens})=1.1$. In all three panels, we show in orange the projected enclosed masses $M_{\rm 2D}(<R_{\rm pivot})$ and deflection-based projected density slopes $\Gamma_{\rm 2D}(R_{\rm pivot})$ of subhalos assigned a collapsed profile from~\cite{Slone_2026} (Section~\ref{sec:subhalo_den_profiles}). For comparison, we show in purple the corresponding projected structures of subhalos assigned the collapsed profile from Figure~1 of~\cite{gurian_may_deep_col} at the final simulation snapshot $t/t_0=437$, where $t_0$ is the characteristic scattering timescale of the initial NFW halo; under the collapse-timescale convention of Eq.~\eqref{eq:t_c_original}, this corresponds to $\tau(t)\simeq12.2$. In both cases, we apply the fiducial tidal truncation prescription described in Section~\ref{sec:subhalo_den_profiles}. The black posterior regions correspond to the inferred perturbers in PJ011646, SDSS J0946+1006, and JVAS B1938+666. The choice of collapsed density profile primarily affects the distributions of higher-mass subhalos at smaller pivot radii, since the two profiles differ significantly at radii $\lesssim0.01r_s$, as shown in Figure~\ref{fig:density_profiles}. For example, at the pivot radius of 90~pc in panel (b), the higher-mass subhalos assigned the profile from~\cite{Slone_2026} exhibit shallower projected density slopes than those assigned the profile from~\cite{gurian_may_deep_col}. However, none of the inferred perturbers discussed in Section~\ref{sec:results:perturbers} have a sufficiently large projected enclosed mass at a sufficiently small pivot radius for these differences to substantially affect our results. We therefore find that our conclusions from Section~\ref{sec:results:perturbers} are robust to either choice of collapsed density profile.

\section{Comparison of \textsc{SatGen} and \textsc{pyHalo} CDM predictions}
\label{app:satgen_pyhalo_comparison}

In this appendix, we generate CDM subhalo realizations of host halos with $M_{\rm vir}(z_{\rm lens})=10^{13}~M_\odot$ at $z_{\rm lens}=0.5$ using \textsc{SatGen}~\cite{satgen_1,satgen_2} and \textsc{pyHalo}~\cite{Gilman_2019_wdm_chills} to compare their lensing-relevant predictions. In Section~\ref{app:projected_shmf}, we show that the tidally evolved, projected subhalo mass function is consistent between CDM realizations generated using both frameworks, and in Section~\ref{app:subhalo_convergence}, we show that their distributions of subhalo lensing convergence are consistent within a circular subhalo-centric aperture with radius 10 milliarcseconds (mas). While these lensing-relevant quantities are not used in this paper for comparison with strong-lens perturbers, they provide a consistency check between the two distinct modeling frameworks. 

\subsection{Projected subhalo mass function}
\label{app:projected_shmf}

Strong-lensing analyses commonly describe the abundance of subhalos near the Einstein radius of a strong-lens system using a projected subhalo mass function (SHMF)~\citep{Gilman_2019_wdm_chills, gilman_jwst4, Nadler_2021_strong_mw_sat}. We adopt the parameterization
\begin{equation}
\frac{d^{2}N_{\rm sub}}{dA~dm_{\rm sub}}=\Sigma_{\rm sub}\frac{m_{\rm sub}^{\gamma}}{m_{\rm pivot,sub}^{~\gamma+1}},
\label{eq:shmf_gamma}
\end{equation}
where $\Sigma_{\rm sub}$ is the projected SHMF normalization, $\gamma$ is the power-law slope, and we set $m_{\rm pivot,sub}=10^{8}~\rm M_{\odot}$. With the definition $\alpha\equiv\gamma+1$, Eq.~\eqref{eq:shmf_gamma} can be written as
\begin{equation}
\frac{d^{2}N_{\rm sub}}{dA~d\ln m_{\rm sub}}=\Sigma_{\rm sub}
\left(\frac{m_{\rm sub}}{m_{\rm pivot,sub}}\right)^{\alpha}.
\label{eq:shmf_alpha}
\end{equation}
We infer $\Sigma_{\rm sub}$ and $\alpha$ from the number of CDM subhalos in the two adjacent mass bins $[10^{7},10^{8}]~\rm M_{\odot}$ and
$[10^{8},10^{9}]~\rm M_{\odot}$. For a circular aperture with area $A$, the expected number of subhalos in mass bin $i$ is
\begin{equation}
\mu_i= A\Sigma_{\rm sub} \int_{\ln m_{{\rm lo},i}}^{\ln m_{{\rm hi},i}}
\left(\frac{m}{m_{\rm pivot,sub}}\right)^{\alpha} d\ln m \equiv
A\Sigma_{\rm sub}J_i(\alpha),
\label{eq:mu}
\end{equation}
where
\begin{equation}
J(m_{\rm lo},m_{\rm hi};\alpha)=
\begin{cases}
\displaystyle
\frac{(m_{\rm hi}/m_{\rm pivot,sub})^{\alpha}-
(m_{\rm lo}/m_{\rm pivot,sub})^{\alpha}}{\alpha},
& \alpha\neq 0,\\
\displaystyle
\ln\!\left(\frac{m_{\rm hi}}{m_{\rm lo}}\right),
& \alpha=0.
\end{cases}
\label{eq:J_alpha}
\end{equation}
Assuming that the counts in the two mass bins are independently Poisson distributed, maximizing their joint likelihood gives
\begin{equation}
\frac{J(m_2,m_3;\hat{\alpha})}{J(m_1,m_2;\hat{\alpha})}=\frac{c_2}{c_1},
\label{eq:alpha_mle}
\end{equation}
and
\begin{equation}
\hat{\Sigma}_{\rm sub}=\frac{c_1+c_2}{A~J(m_1,m_3;\hat{\alpha})},
\label{eq:sigma_mle}
\end{equation}
where $c_1$ and $c_2$ are the subhalo counts in the two mass bins.

\begin{figure}[t]
    \centering
    \includegraphics{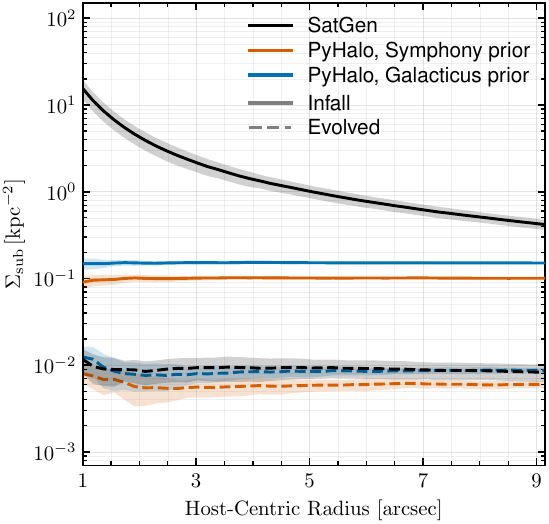}
    \caption{
    Projected subhalo mass function normalization (Eq.~\eqref{eq:sigma_mle}) versus host-centric circular aperture radius for CDM subhalos generated using \textsc{SatGen} and \textsc{pyHalo}, averaged over 24 subhalo realizations of host halos with $M_{\rm vir}(z_{\rm lens})=10^{13}\rm~M_\odot$ at $z_{\rm lens}=0.5$. Solid lines use subhalo infall masses, while dashed lines use evolved bound masses at the lens redshift. The \textsc{pyHalo} curves correspond to abundance calibrations based on either the \textsc{Symphony}~\cite{Nadler_2023_symphony} cosmological simulations (orange) or the \textsc{Galacticus}~\cite{Benson_2012_galacticus} semi-analytic model (blue). Shaded bands show the $1\sigma$ realization scatter.
    }
    \label{fig:sigma_sub_satgen_vs_pyhalo_appendix}
\end{figure}

We estimate $\Sigma_{\rm sub}$ from 24 CDM subhalo realizations generated using \textsc{SatGen} and \textsc{pyHalo}. Figure~\ref{fig:sigma_sub_satgen_vs_pyhalo_appendix} shows $\Sigma_{\rm sub}$ as a function of host-centric circular aperture radius. For each framework, we estimate $\Sigma_{\rm sub}$ using both the subhalo masses at infall and the evolved bound masses at the lens redshift. Using first-order error propagation, we verified that the uncertainty associated with the two-bin fit is negligible compared to the scatter across realizations for the aperture radii considered. The weak dependence of the inferred normalization on aperture radius indicates that the projected subhalo number density is approximately uniform over the radial range considered, consistent with theoretical expectations~\citep{gannon_2025_dm_lensing}. 

The SHMF normalizations inferred using infall masses from \textsc{SatGen} and \textsc{pyHalo} are not expected to coincide because the two frameworks generate subhalo populations differently. \textsc{SatGen} explicitly generates merger trees and tidally evolves individual subhalos from infall to the lens redshift, as described in Sec.~\ref{sec:satgen_real_in_cdm}. In contrast, \textsc{pyHalo} generates subhalo populations at the lens redshift without following individual orbital histories. Infall masses are sampled in projection using Eq.~\eqref{eq:shmf_gamma}, whose normalization $\Sigma_{\rm sub}$ is specified as an input parameter~\citep{gilman_jwst4}. Bound mass fractions are sampled from the tidal-evolution model of~\cite{du_2025_faster_sam}, which is calibrated using \textsc{Galacticus}~\cite{Benson_2012_galacticus} and implemented in \textsc{pyHalo}. Tidal tracks then map the sampled bound mass fraction onto rescaled and truncated NFW profiles at the lens redshift. Following~\cite{gilman_jwst4}, we adopt the values of $\Sigma_{\rm sub}$ for their \textsc{Galacticus}~\cite{Benson_2012_galacticus} and \textsc{Symphony}~\cite{Nadler_2023_symphony} abundance priors, $\Sigma_{\rm sub}=0.15~{\rm kpc}^{-2}$ and $\Sigma_{\rm sub}=0.10~{\rm kpc}^{-2}$, corresponding to the blue and orange solid lines in Figure~\ref{fig:sigma_sub_satgen_vs_pyhalo_appendix}, respectively. After applying the tidal-evolution prescription in \textsc{pyHalo}, we use Eq.~\eqref{eq:sigma_mle} to infer the corresponding evolved normalizations, shown by the dashed blue and orange curves. Despite \textsc{SatGen} and \textsc{pyHalo} using different tidal evolution prescriptions, these frameworks predict consistent evolved SHMF normalizations, and hence consistent projected subhalo abundances over the aperture radii relevant for group-scale strong lenses. 

\subsection{Distribution of subhalo convergence}
\label{app:subhalo_convergence}

We compare the lensing strength of CDM subhalos between \textsc{SatGen} and \textsc{pyHalo} by considering their mean convergence within a 10-milliarcsecond subhalo-centric aperture, which is representative of the angular impact parameter over which low-mass halos produce several-percent perturbations to the magnifications of lensed-quasar images~\citep{nierenberg_2023_jwst1_10mas, gilman_jwst4}. For a circularly symmetric subhalo density profile, the convergence profile is defined as
\begin{equation}
\kappa(\boldsymbol{\theta})=
\frac{\Sigma(\boldsymbol{\theta})}{\Sigma_{\rm crit}},
\end{equation}
where $\Sigma(\boldsymbol{\theta})$ is the projected surface mass density profile and $\Sigma_{\rm crit}$ is the critical surface density for lensing~\citep{saha_2024_strong_lens_essentials}. The integrated convergence within a circular subhalo-centric aperture with radius $R$ is
\begin{equation}
K(<R)=\int_{|\boldsymbol{\theta}|<R}
\kappa(\boldsymbol{\theta})~d^{2}\boldsymbol{\theta}=2\pi\int_{0}^{R}\kappa(\theta)
~\theta~d\theta.
\label{eq:integrated_subhalo_convergence}
\end{equation}
The aperture-averaged convergence is then
\begin{equation}
\left\langle\kappa\right\rangle_{<R}\equiv\frac{K(<R)}{\pi R^{2}}=\frac{2}{R^{2}}\int_{0}^{R}\kappa(\theta)~\theta~d\theta,
\label{eq:average_subhalo_convergence}
\end{equation}
and the corresponding magnitude of the deflection angle $\alpha$ at radius $R$ satisfies
\begin{equation}
\alpha(R)=\frac{K(<R)}{\pi R}=R\left\langle\kappa\right\rangle_{<R},
\label{eq:deflection_average_convergence}
\end{equation}
for an axisymmetric lens~\citep{saha_2024_strong_lens_essentials}. Thus, the aperture-averaged convergence $\langle\kappa\rangle_{<R}$ determines the deflection angle produced by the subhalo mass profile, which is closely related to the localized image distortions generated by dark matter subhalos inferred in strong-lensing analyses~\citep{Minor_2017_inference,sengul_2022_inference,vegetti_2009_inference}.

\begin{figure}[t]
    \centering
    \includegraphics{
    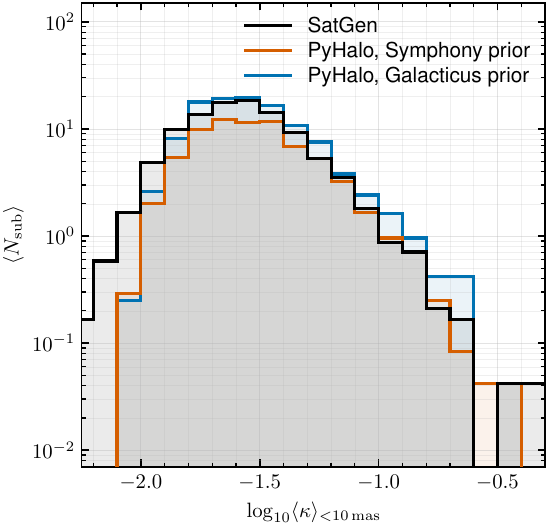}
    \caption{
     Histograms of aperture-averaged convergence (Eq.~\eqref{eq:average_subhalo_convergence}) within a circular $10~{\rm mas}$ subhalo-centric aperture of CDM subhalos generated using \textsc{SatGen} and \textsc{pyHalo} under \textsc{Symphony} or \textsc{Galacticus} model priors. We use 24 subhalo realizations of host halos with $M_{\rm vir}(z_{\rm lens})=10^{13}~\rm M_{\odot}$ at lens redshift $z_{\rm lens}=0.5$ and source redshift $z_{\rm source}=1.5$. $\left\langle N_{{\rm sub}}\right\rangle$ is the realization-averaged number of subhalos in log-convergence bins of width $0.1$ dex. At the lens redshift, only subhalos with evolved bound masses $m_{\rm bound}\geq10^{7}~\mathrm{M}_{\odot}$ that lie within a circular projected host-centric aperture of $3.3\rm~arcsec$ are considered.
    }
    \label{fig:avg_convergence_satgen_vs_pyhalo_appendix}
\end{figure} 

Figure~\ref{fig:avg_convergence_satgen_vs_pyhalo_appendix} shows distributions of aperture-averaged convergence for CDM subhalos generated using \textsc{SatGen} and \textsc{pyHalo} under both \textsc{Symphony} and \textsc{Galacticus} priors described in Sec.~\ref{app:projected_shmf}. \textsc{SatGen} CDM subhalos are modeled using the density profile in Eq.~\eqref{eq:green_profile}, which is implemented in \textsc{lenstronomy}~\citep{birrer_2018_Lens1, Birrer_2021_Lens2}. We use this profile to construct their convergence profiles and calculate $\langle\kappa\rangle_{<10~{\rm mas}}$ using Eq.~\eqref{eq:average_subhalo_convergence} with $R=10~{\rm mas}$. \textsc{pyHalo} CDM subhalos are modeled with rescaled truncated-NFW profiles described in~\cite{gilman_jwst4}, whose parameters are used to construct the corresponding truncated-NFW profile using \textsc{lenstronomy}. We then analogously calculate $\langle\kappa\rangle_{<10~{\rm mas}}$ for each subhalo. We use 24 subhalo realizations of host halos with $M_{\rm vir}(z_{\rm lens})=10^{13}~\rm M_{\odot}$ at lens redshift $z_{\rm lens}=0.5$ and source redshift $z_{\rm source}=1.5$, over which the counts in each convergence bin are averaged. We only use evolved subhalos with bound masses $m_{\rm bound}\geq10^{7}~\mathrm{M}_{\odot}$, whose projected positions lie within a circular 3.3~arcsec host-centric aperture. \textsc{SatGen} and \textsc{pyHalo} predict approximately consistent distributions of subhalo convergence despite using different tidal evolution prescriptions, and hence both frameworks predict comparable distributions of subhalo deflection angles at $10~{\rm mas}$. 

\section{Summary of PJ011646, SDSS J0946+1006, and JVAS B1938+666}
\label{app:observed_perturbers}

We summarize the observational history and inferred properties of the three candidate strong-lens perturbers in PJ011646, SDSS J0946+1006, and JVAS B1938+666, emphasizing the gravitational imaging analyses from which we adopt their projected enclosed mass and density slope posteriors. PJ011646 is a strong-lens system with $z_{\rm lens}=0.555$~\cite{Liu_2024_PJ011646_z_lens} and $z_{\rm source}=2.125$ identified as part of the PASSAGES sample of Planck-selected sources~\citep{PJ011646_discover_passages}. Using high-resolution $\sim0.1''$ ALMA Band~7 dust-continuum observations, the detection of a potential dark matter perturber was reported in~\cite{PJ011646_2026_dir_det_original}. The interferometric data were modeled in the $uv$ plane using a smooth elliptical power-law (EPL) mass distribution with external shear and third- and fourth-order multipoles. A grid-based search in the image plane for a spherical NFW halo identified a candidate, and the model including the perturber was preferred over the fiducial smooth macromodel by $\Delta\ln Z\simeq22.5$, corresponding to a detection significance of approximately $5.8\sigma$. Assuming that the perturber lies at the lens redshift, the analysis reported $M_{200}=2.78^{+0.43}_{-0.66}\times10^{10}~\mathrm{M_\odot}$ and $c_{200}=30^{+5}_{-7}$, with a concentration that they argued is consistent with a typically tidally stripped CDM subhalo (see Section~\ref{sec:PJ011646} for our interpretation). The projected enclosed mass was most robustly constrained at an angular pivot radius of $0.29''$, within which they inferred $M_{\rm 2D}=3.57^{+0.16}_{-0.14}\times10^9~\mathrm{M_\odot}$.

SDSS J0946+1006, also known as the ``Jackpot'' lens, is a strong-lens system with $z_{\rm lens}=0.222$ and three lensed background sources. The system was originally observed as a double Einstein ring in HST/ACS imaging with an inner source at redshift $z_{s1}=0.609$~\cite{Gavazzi_2008_j0946_discovery}. The second source redshift was subsequently measured to be $z_{s2}=2.035$ using VLT/X-Shooter and MUSE observations~\cite{smith_collett_2021_j0946_second_source}, while a third multiply imaged source at $z_{s3}=5.975$ was discovered using VLT/MUSE data~\cite{collett_2020_j0946_third_source}. The first gravitational-imaging detection of a dark perturber in this system was reported by~\cite{Vegetti_2010_dir_det_original} using HST/ACS F814W data. Their non-parametric reconstruction identified a localized perturbation near the inner Einstein ring, which they modeled with a tidally truncated pseudo-Jaffe profile. The analysis inferred a perturber mass of $(3.51\pm0.15)\times10^9~\mathrm{M_\odot}$ with no luminous counterpart detected in the image. The perturber was then analyzed in several subsequent studies using various lens and source models~\cite{Minor_2021_J0946_concentration,Nightingale_2024_subhalo_scan,Enzi_2024_cdm_unexpect,Despali_2025_cdm_unexpect,Tajalli_2025_J0946_sidm,Ballard_2024_J0946}. However, its inferred structural properties remain sensitive to modeling assumptions such as macromodel flexibility and source regularization~\cite{Tajalli_2025_J0946_sidm,Ballard_2024_J0946}, and whether the perturber is assumed to be completely dark or allowed to host faint stellar luminosity~\cite{He_2025_not_so_dark}. Subsequent analyses have investigated its consistency with CDM predictions and possible SIDM interpretations~\citep{Enzi_2024_cdm_unexpect,Despali_2025_cdm_unexpect,Tajalli_2025_J0946_sidm,Li_2025_dark_dot}. The HST F814W data were reanalyzed by~\cite{minor_2025_direct_detect} using image-pixel supersampling and double-source-plane modeling. The analysis recovered a detection that is substantially less sensitive to the source-regularization prior, obtaining a log-Bayes factor $\Delta\ln E\simeq143$, corresponding approximately to a $17\sigma$ detection. At $95\%$ confidence, the projected enclosed mass within $1~\mathrm{kpc}$ was constrained to $(2.2-3.4)\times10^9~\mathrm{M_\odot}$, while the average logarithmic slope of the local surface density, $d\ln\Sigma/d\ln R$, over $0.75$--$1.25~\mathrm{kpc}$ was constrained to be steeper than $-1.75$.

JVAS B1938+666 is a radio and near-infrared strong-lens system with lens redshift $z_{\rm lens}=0.881$ and background source redshift $z_{\rm source}=2.059$~\citep{Riechers_2011_jvasb_source_z,Tonry_Kochanek_2000_JVASB,King_1997_JVASB,King_1998_JVASB}. Using Keck-II/NIRC2 adaptive-optics observations, the detection of a $\sim1.9\times10^8~\mathrm{M_\odot}$ perturber was reported in~\cite{Vegetti_2012_dir_det_original} with approximately $12\sigma$ confidence. This object, commonly referred to as detection $\mathcal{A}$, was subsequently inferred to be a background halo at redshift $1.42^{+0.10}_{-0.15}$~\citep{sengul_JVASB_los_halo}. More recently, however, the Keck adaptive-optics data were reanalyzed in~\cite{Tajalli_2025_J0946_sidm} and detection $\mathcal{A}$ was found to be a foreground halo at redshift $0.13\pm0.07$. A second substantially lower-mass $\sim10^6~\rm M_{\odot}$ perturber, commonly called detection $\mathcal{V}$, was claimed to be discovered by~\cite{Powell_2025_dir_det_original} using $1.7~\mathrm{GHz}$ VLBI observations with an angular resolution of approximately $5~\mathrm{mas}$. It was found along an extremely thin $\sim200~\mathrm{mas}$-long lensed arc using a non-parametric approach, while independent parametric modeling confirmed detection $\mathcal{V}$ with approximately $26\sigma$ confidence and measured a projected enclosed mass $M_{\rm 2D}(<80~{\rm pc})=(1.13\pm0.04)\times10^6~\mathrm{M_\odot}$~\citep{Powell_2025_dir_det_original}. In~\cite{Vegetti_2026_cdm_unexpect}, detection $\mathcal{V}$ was analyzed using a broad set of density profiles and perturber-redshift assumptions. Although its projected density slope is sensitive to the assumed profile, the projected enclosed mass is most robustly constrained at $\simeq90~\mathrm{pc}$. They found that single-component NFW models require concentrations substantially larger than CDM expectations, and discussed whether the compact structure could instead be a core-collapsed SIDM halo, an ultra-compact dwarf galaxy containing a central black hole, or a nuclear star cluster.

\bibliographystyle{JHEP}
\bibliography{ref}

\end{document}